# A systematic literature review on the application of analytical approaches and mathematical programming in public bus transit network design and operations planning: Part II

**Reza Mahmoudi**

Department of Civil Engineering,

University of Calgary,

AB T2N 1N4, Calgary, Alberta, Canada

E-mail: reza.mahmoudi@ucalgary.ca

**Saeid Saidi**

Department of Civil Engineering,

University of Calgary,

AB T2N 1N4, Calgary, Alberta, Canada

E-mail: ssaidi@ucalgary.ca

**S. Chan Wirasinghe**

Department of Civil Engineering,

University of Calgary,

AB T2N 1N4, Calgary, Alberta, Canada

E-mail: wirasing@ucalgary.ca

# Attention

Please note that this literature review encompasses four distinct papers, two of which are published in peer-reviewed journals and two are available on arXiv.org. To optimize your reading, please read the papers in the following order:

1) "A Systematic Literature Review on the Application of Analytical Approaches and Mathematical Programming in Public Bus Transit Network Design and Operations Planning – Part I" (published online on arXiv.org)
2) Mahmoudi, R., Saidi, S. and Wirasinghe, S.C., 2024. A critical review of analytical approaches in public bus transit network design and operations planning with focus on emerging technologies and sustainability. Journal of Public Transportation, 26, p.100100. (https://www.sciencedirect.com/science/article/pii/S1077291X24000201)
3) "A Systematic Literature Review on the Application of Analytical Approaches and Mathematical Programming in Public Bus Transit Network Design and Operations Planning – Part II" (published online on arXiv.org)
4) Mahmoudi, R., Saidi, S. and Emrouznejad, A., 2025. Mathematical Programming in Public Bus Transit Design and Operations: Emerging Technologies and Sustainability–A Review. Socio-Economic Planning Sciences, p.102155. (https://www.sciencedirect.com/science/article/pii/S0038012125000047)

***Note.*** *For any inquiries or contributions to the unpublished papers, please contact the corresponding author or other authors.*

***Corresponding Author:*** *Reza Mahmoudi, Postdoctoral Fellow, University of Toronto, reza.mahmoudi@ucalgary.ca, reza.mahmoudi@utoronto.ca, j.mahmoudi.reza@gmail.com*

# A systematic literature review on the application of analytical approaches and mathematical programming in public bus transit network design and operations planning: Part II

## Abstract

Among all public transit modes, bus transit systems stand out as the most prevalent and popular. This prominence has spurred a significant body of research addressing various aspects of bus systems. In the literature, analytical approaches and mathematical programming are predominantly used to explore the Public Bus Transit Network Design Problem and Operations Planning (PBTNDP&OP). Part I of our study presented statistical analyses of literature applying these methodologies to PBTNDP&OP, along with a comprehensive review of analytical papers, highlighting the strengths and weaknesses of each approach. In Part II, we delve into the applications of mathematical programming within PBTNDP&OP, building upon the 15 major sub-categories identified in Part I. We have critically analyzed the identified papers within these sub-categories from various perspectives, including the problems investigated, modeling methods employed, decision variables, network structures, and key findings. This critical review highlights selected papers in each category. Finally, acknowledging existing research gaps, we propose potential extensions for future research. Despite the extensive array of publications, numerous topics still warrant further exploration. Notably, sustainable PBTNDP&OP and challenges associated with integrating emerging technologies are poised to dominate future research agendas.

## 1. Introduction

Among all public transportation services (PTSs), undoubtedly, public bus transit service (PBTS) is the most prevalent PTS all around the world. PBTS, with its different versions (regular buses, BRT, feeder, shuttle, school bus, etc.), is applicable in any city with any transportation network size/structure and any population density. Because of this popularity and applicability, the public bus transit network design problem and operations planning (PBTNDP&OP) have attracted researchers' and transportation managers' attention.

From the mathematical perspective, PBTNDP&OP is assumed to be one of the most complicated transportation problems to analyze (Ibarra-Rojas et al., 2015). The multiplicity of the beneficiaries, decision-makers, and evaluation criteria, interdependencies among the solutions, uncertainty, and missing data are only some of the numerous reasons for the high complexity of PBTNDP&OP. Different methods have been applied in the literature to study these problems and find optimal or near optimal solutions. While some researchers have applied analytical approaches, others have employed the advantages of mathematical programming and heuristic/metaheuristic algorithms. Obviously, all of these studies have helped the public transportation industry. However, it would be an interesting challenge to conduct a comprehensive literature review on the existing studies belonging to the mentioned groups.

In part I, we presented some statistical analyses of the existing literature on the applications of analytical approaches and mathematical programming in the PBTNDP&OP. We also compared the advantages and disadvantages of each methodology against the other one. Then, we reviewed the literature on the applications of analytical approaches in the PBTNDP&OP. In Part II, we will review the existing literature on the applications of mathematical programming in the PBTNDP&OP, by considering published studies between 1968 and 2021. To this end, we will consider 15 major identified subtopics in the literature of the applications of mathematical programming in the PBTNDP&OP and will review the existing literature related to each subtopic. In particular, this study aims to analyze the existing literature and review the definitions, classifications, objectives, constraints, network topology decision variables, modeling approaches, and key findings related to the PBTNDP&OP. The objectives are twofold: first, presenting a comprehensive review and criticism of the selected papers in each category (to find out how the papers have been selected, see Part I); second, identifying recent trends in the conducted studies, existing research gaps, and possible extensions for future research.

The rest of the paper has been organized as follows: A comprehensive review, including reviewing understudied problems, applied methods, network structures, and findings, has been presented in section 2, "Comprehensive literature review". The identified research gaps and possible extensions have been discussed in section 3. Finally, "Conclusion" includes the concluding remarks.

## 2. Comprehensive Literature Review

In Part I, 15 major sub-categories were identified in the related literature to the applications of mathematical programming in PBTNDP&OP. The identified major subcategories are: 1) Accessibility and coverage, 2) Feeder transit network design, 3) Service pricing and fare management, 4) Fleet management, 5) Dispatching policy, 6) Stop spacing, 7) Mode split and traffic assignment, 8) Network design and dispatching policy, 9) Network design and routing, 10) Green and sustainable public bus network design, 11) Flexible and fixed transit service, 12) Travel time and reliability, 13) Disaster and disruption management, 14) Bus priority, and 15) Emerging technologies and electric buses.

Following, the existing literature in each sub-category has been reviewed. Please note that the related literature to the applications of analytical approaches and mathematical programming in "Emerging technologies in public bus transit systems" and "Sustainability in PBTNDP&OP" will not be reviewed in this paper, as the related papers to these concept has been already published and are available in Mahmoudi et al. (2024) and Mahmoudi et al. (2025).

### 2.1. Accessibility and coverage

There are many different criteria in the literature used to study PBTNDP&OPs. Among all of these criteria, operation costs, waiting costs, and in-vehicle costs have been used more often than others, where the objective was to minimize the summation of these costs. However, it can be seen in the literature that accessibility and coverage are other criteria that are being considered by researchers more than before. Maximizing spatial accessibility and coverage is one of the main objectives of any public transit service, which is to serve as many passengers as possible. Accessibility can be defined as the geographical coverage and access of residents to public transit services.

While adding more stops to a bus route can lead to greater access, it will result in slower transit service, which is not desirable for both users and the agency. Increasing access means lower walking costs to and from the stops, but longer travel times mean lower convenience (for more passengers) and a drop in ridership. In addition, with the same resource (total bus operating hours being constant), higher number of stops, and considering a travel time constraint, more stops (equivalently higher travel time) will decrease the service area (equivalently less geographic coverage). Considering this conflict, Murray and Wu (2003) discussed different models to consider accessibility in public bus transit services and analyzed the benefits of these models by applying them to the real case of Columbus, Ohio.

Murray (2003) proposed a model to simultaneously increase service area (coverage) and increase users' access to public transit services. Then, integrating this model with a GIS environment, they applied it to the case of public transit service in Brisbane, Australia, in order to make strategic plans. Considering the trade-off between service quality and access in public bus transit service, Wu and Murray (2005) developed a multiple-route optimization model that is able to identify redundant or underutilized stops to be removed from the bus routes. The objective was to enhance service quality. The proposed model was a maximal covering/shortest-path model. Zhao and Zeng (2006) developed a stochastic model to design an optimal large scale transit network with the minimum number of transfers. In order to maximize coverage, the proposed model was used to find reasonable route directness. To solve the proposed model, they integrated Simulated annealing (SA) and GA as two metaheuristic algorithms. Aboolian et al. (2016) conducted another study on maximizing users' access to public transit services. First, they proposed an analytical method to find the optimal features of the facilities (number, location, etc.) in order to design a public transit network with maximum accessibility. They defined the accessibility level to a facility as the average time needed to receive a service from that facility. Then, they developed a nonlinear integer programming model to take the congestion at the facilities and the customer-choice environment into account in evaluating service quality and used ε-optimal algorithm to find a solution for this model.

Marseglia et al. (2019) proposed a mathematical model to find the optimal set of stations/stops in a bus corridor/route leading to the maximum population covered by the public transit system by avoiding the zigzag effect (e.g. zigzag roads). In order to take the zigzag effect under control, they introduced the concept of a curvature concept for polygonal lines and considered maximizing the population covered by the public transit system and minimizing curvature as the objectives in the proposed method. In order to increase the access of seniors to public bus transit services and considering two different schemes, Chen et al. (2020) proposed a bi-level model to redesign the bus transit routes in a network. The objective of the proposed model was to identify optimal stop locations and routes, considering the shortest distance as the evaluation criterion. In the upper level, the seniors' access was being maximized. An exact algorithm for the first scheme and a GA for the second were proposed to find optimal and near to optimal solutions. Chen et al. (2021b) developed a mathematical program to identify optimal stop location, route design, and dispatching policy for a customized bus transit service. The objectives of the proposed model were to maximize users' accessibility and minimize operating costs.

Table 1 represents the main features of the reviewed papers in this subsection. According to this table, max. of ridership, max. of accessibility, and max. of service coverage are the three most commonly considered objective functions in this section, while route location and stop location are main decision variables. The authors have tested different algorithms to solve the proposed models in this section; however, GA is the most applied algorithm. Satisfaction and equity are other objectives that have been considered by Camporeale et al. (2016). These objectives are excellent criteria to reflect the coverage and accessibility of the PBTSs. However, as these criteria are mostly considered as qualitative social criteria, a challenging issue will arise when it comes to quantifying them. Developing different indices to quantify criteria, in particular social criteria, can be an interesting contribution to the literature of coverage and accessibility in PBTSs.

**Table 1.** Main features of the reviewed papers in subcategory 1.

| Reference | Objective Function | Constraints | Transit mode | Network structure | Demand Pattern | Decision Variables | Solution Approach |
|---|---|---|---|---|---|---|---|
| Murray and Wu (2003) | Max. total potential ridership | Number of transit stops | Bus | Real world | Given O-D matrix | Stop location | CPLEX |
| Murray (2003) | Max. total potential ridership (coverage) | Number of transit stops | Bus | Real world | Given O-D matrix | Stop location | CPLEX |
| Wu and Murray (2005) | Min. Total travel time, Max. total potential ridership | Demand-flow | Bus | Real world | Given O-D matrix | Stop location | SAW & CPLEX |
| Zhao and Zeng (2006) | Max. of service coverage | Route length & direction | Bus | Real world | Given O-D matrix | Transit route | SA & GA |
| Aboolian et al. (2016) | Max. accessibility | Number of served passengers and population of an area & service capacity | General | Real world | Elastic demand | The number, location and capacity of facilities | ε-optimal |
| Camporeale et al. (2016) | Min. total user & agency cost / max. satisfaction (equity) | Car equity, bus equity, demand-flow & cycle time | Multi-modal | Small idealized network | Given O-D matrix | Signal setting & flow | GA |
| Marseglia et al. (2019) | Max. population covered by the system and max. curvature | Max. and min of line length, demand-flow, route directions & curvature | Bus | Real world | Given O-D matrix | Alignment | Greedy Constructive algorithm |
| Chen et al. (2020) | Min distance / Max. accessibility | Maximum tolerable route length, between stop spacing, | Bus | Grid network | Parameter | Transit routes and stop location | Exact & GA |
| Chen et al. (2021b) | Max. Accessibility & min. operating cost | User assignment, vehicle running time, Demand-flow, vehicle capacity, fleet size, route distance & number of stops | Bus | Real world | Parameter | Transit route, timetable & stop location | GA & VNS |

## 2.2. Feeder transit network design

There are a good number of studies that have applied mathematical programming and algorithms to study the feeder bus transit network design problem (FBNDP). Martins and Pato (1998) investigated the feeder bus network design problem by developing an optimization approach considering the total users' and agency's costs as the objective function to be minimized along with resource constraints and travel demands. The transit routes and associated service frequencies were the main outcomes of the proposed model. They assumed that the feeder bus service would feed a rail transit system in a city. To find acceptable solutions, they used Tabu search heuristics. Chien et al. (2001) conducted another similar study. They proposed a model with the same objective function and main decision variables to investigate the FBNDP. To this end, they considered different constraints related to the available budget, irregular grid street patterns, delays in the intersections, a nonhomogeneous demand pattern in the network, etc. They used a GA to find a near to optimal solution for the proposed model. Kuan et al. (2004) designed two metaheuristic algorithms to find near to optimal solutions for the FBNDP based on the concept of SA and Tabu search. They applied the proposed algorithms to solve different numerical examples and compared their performances from different perspectives. Considering a feeder bus service that is feeding a rail transit service, Verma and Dhingra (2005) proposed a model to solve the FBTNDP. They considered the case that the city needs to have a new mass transit service, such as a rail system. The proposed approach was used to identify an initial set of shortest paths as candidate feeder bus routes and then to generate K shortest paths for any origin-destination pairs (between the stations and feeder transit terminals). Finally, they applied a GA to choose one route among all the identified possible candidates. In order to solve FBTNDP, Kuan et al. (2006) developed two metaheuristics based on the concepts of GA and ant colony (ACO) optimization. They applied the developed algorithms to solve different numerical examples and analyzed the performance of the algorithms from different perspectives.

Shrivastava and O'Mahony (2006) developed a GA to solve the routing problem for a feeder service and obtain service frequency for each route. Their objective was to find a feeder service frequency that results in schedule coordination between feeder and trunk services. To this end, they assumed that the service frequency of the trunk service is given and the decision variable is the frequency of the feeder service. Another similar study has been conducted by Shrivastava and O'Mahony (2009). They applied a hybrid GA-a heuristic algorithm to solve the same problem discussed in the previous study and compared their performance. The results showed that their algorithm obtained better results compared to the previous study. Lin and Wong (2014) developed a multi-objective programming to investigate FBNDP considering minimizing length of the routes, minimizing maximum travel time on the routes, and maximizing coverage as the objective functions. In order to take the uncertainty of parameters into account, they used fuzzy numbers theory. Pan et al. (2015) proposed a multilevel mixed integer optimization model to investigate the problem of optimal zone and route design for a flexible feeder bus service. They assumed the feeder system would serve passengers in an irregular-shaped transit network, and the fleet size was given. Also, the travel time between any successive demand points was assumed to be a known value. They suggested a heuristic to find a near to optimal solution for the proposed model within an acceptable execution time. Almasi et al. (2015) applied three different metaheuristic algorithms, including GA, PSO, and imperialist competitive algorithm, to find near to optimal design of feeder transit services and coordinate associated service schedules. They applied the algorithms to solve different cases and numerical examples and analyzed the results.

Gschwender et al. (2016) considered a feeder-trunk transit structure as a superior structure to serve the passengers in a city and compared it with three other different linear structures where there isn't any feeder transit service. Considering both users' and the agency's costs, for each structure, they found the optimal fleet size and bus size. Their results showed that the optimal structure will depend on demand patterns and the trunk corridor's length. Also, transfer costs will play another pivotal role in the process of selecting the optimal structure. Considering different factors, Zhu et al. (2017) first developed a demand prediction model for the case of adding a feeder bus service to the network, then used a logit model to predict the users' flow distributions. They generated routes for feeder services using a circular route model and used a GA to solve the problem. Li et al. (2018) proposed a mixed integer programming to model the problem of optimal dispatching policy for a flexible feeder bus transit service considering both users' and the agency's costs. The decision variables were choosing pick-up locations and transit routes. They used commercial solvers to solve the proposed model for small scale networks and a GA to solve it for large scale networks. Sun et al. (2018) considered a network including a rail transit service that is being fed by a flexible route bus transit service, and then, proposed a mixed integer optimization model to assign each vehicle to respond to a set of demands and transfer the user to the station. The objective function of the proposed model was to minimize the total agency's costs and maximize the satisfaction of the users. They assumed that each user would consider a time window for its pickup time and in-vehicle time. To find near to optimal solutions, they also developed a Bat algorithm to solve the proposed model. Huang et al. (2018) considered a bi-modal transit service where a bus service feeds a rail transit service, newly added to the network. They also assumed there is a bus service serving passengers independently (regular/non-feeder bus transit service). Then, they proposed a model to address public transit network design and frequency setting problems for each mode. They used a cluster-based approach to find hubs among rail stations, a heuristic algorithm to find lines for the independent bus services, and the travelling salesman problem to find the feeder bus service routes. Finally, they developed a bi-level optimization model to obtain service frequencies for all modes and used the artificial bee colony (ABC) algorithm to solve this model.

Sun et al. (2019) proposed a mixed integer programming model to investigate the demand-responsive FBNDP where the feeder system is serving a rail transit service. To capture the uncertainty effects in the users' demand, they used fuzzy numbers theory. The objective of the proposed model was minimizing the total travel distance for all routes, when time windows for pick up and in-vehicle travel time have been considered in the modeling process. Using the credibility theory, they converted the proposed model to a deterministic linear optimization model. They also developed an ACO algorithm to find acceptable solution(s) for the proposed model and compared its performance to that of other algorithms including standard ACO, PSO and GA. Park et al. (2019) developed a method to design a feeder bus transit service to serve demands that are currently being served by taxis during peak hours. To this end, first using real-time GPS data of taxies, Road2vec, and k-means clustering analysis, they found the busiest clusters that are being served by taxies. Then, for each identified cluster, they designed a feeder bus transit route. Finally, an integer programming model was used to obtain the optimal location of the bus stops on the feeder routes.

In order to design a bi-modal coordinated feeder-trunk transit corridor with an optimal total cost, Yang et al. (2020) developed a two-stage programming model. In the first stage, they obtained a feasible network design and an optimal uncoordinated scheme by integrating continuum approximation and a discrete model. In the second stage, stop/station and feeder route locations, the feeder route length, and the distribution of the passenger flows were obtained. To solve the proposed model, they used a nested two-phase optimization algorithm, analytical approaches, and a GA. Wei et al. (2020) proposed an optimization model to find the closest pickup and drop off stops for passengers, origin-to-terminal transit routes, and associated service frequencies for a feeder transit service. The objective function of the proposed model was minimizing the walking, in-vehicle, and waiting costs, simultaneously. As the key results, they obtained the maximum walking distance, the rate of the load factor associated with each route, and the lower and upper bounds for the service frequencies related to each route. To solve the proposed model, they integrated a GA with the Dijkstra's algorithm. Developing a mathematical programming model for multi-level multi-mode PBNTDP, Wang et al. (2020b) designed the skeleton, arterial, and feeder networks in a city. For each level, they also proposed different optimization models depending on the network specifications. Wang et al. (2020c) developed a two stage mathematical programming for responsive FBNDP and service scheduling. The objective function was to minimize summation vehicle operation and users' travel costs, where the constraints were related to the capacity and running time of the vehicles, and time windows for the users. They considered two types of demand for the service: reservations and real-time demand. To solve the problem, they first obtained the value of the decision variables based on the reservation demands (first stage), then dynamically updated these values based on the real-time demand, using quantitative batch treatment. A GA was used to solve the proposed model.

Galarza Montenegro et al. (2021) considered the case of mixed on-demand and traditional feeder bus transit services, assuming that some stops will always be served while others will be served only upon request in the nearby area. To solve the proposed optimization method, they used a large neighborhood search (LNS) algorithm. As a key result, they showed that under the demand-responsive service, the network will perform way better than under traditional one. Li et al. (2021) studied the problem of coordinated feeder bus and high speed rail transit network design problem by considering time windows. They investigated the routing problem for feeder service and passenger assignment problems for the rail service, where the objective was to minimize the average delay time of the users in the rail stations. To find near to optimal solutions, an adaptive LNS was used to deal with solving the proposed model. Wei et al. (2021) developed a bi-level optimization model for FBNDP, considering minimizing in-vehicle time as the objective of the upper level and total users' access cost to be minimized as the objective of the lower level model. The main advantage of the developed approach was the ability to include different real data gathered by a GIS system in the programming. They applied an ACO algorithm to find a solution for the proposed model in a reasonable execution time.

Table 2 represents the main features of the reviewed papers in this subsection. According to this table, minimization of the sum of access costs, waiting costs, in-vehicle costs, and operation costs is the most commonly used objective function. The most commonly considered constraints are related to demand, network geography, fleet capacity, network connectivity, traffic flow, and stop/station locations. Routing and service frequency are the two main decision variables in this section. The authors have tested different algorithms to solve the proposed models in this section; however, GA is the most applied algorithm. Based on Table 2, all reviewed papers have been conducted on bus-rail and bus transit systems. However, it would be interesting to investigate integrating feeder bus transit services with other emerging feeder services such as e-haul, e-bikes, shared vehicles, etc.

**Table 2.** Main features of the reviewed papers in subcategory 2.

| Reference | Objective Function | Constraints | Transit mode | Network structure | Demand Pattern | Decision Variables | Solution Approach |
|---|---|---|---|---|---|---|---|
| Martins and Pato (1998) | Min. sum users and agency cost | Resource, route length and capacity, total fleet mileage & demand constraints | Bus-Rail | Real world | Time dependent given parameter for each stop | Routing and service frequency | Tabu search |
| Chien et al. (2001) | Min. sum access cost, waiting and in-vehicle costs & operation cost | Geographic, capacity & budget constraints | Bus | Rectangular grid network | Heterogeneous demand distribution | Routing and service frequency | GA |
| Kuan et al. (2004) | Min. sum of waiting, in-vehicle & operation cost | Route feasibility, stop-route matching, bus flow in each node, stop-station matching, route capacity, fleet capacity & route length | Bus | Benchmark network | Given for each node | Bus routes & stop location | SA & Tabu search |
| Verma and Dhingra (2005) | Min. in-vehicle travel cost | Max unsatisfied demand | Bus-Rail | Real world | Demand estimation function | Bus route | GA |
| Kuan et al. (2006) | Min. sum of waiting, in-vehicle & operation cost | - | Bus | Test networks | Hourly demand given for each stop | Bus routes & service frequency | SA & Ant colony optimization |
| Shrivastava and O'Mahony (2006) | Min. sum of transfer cost, travel cost and operation cost | Max/min load rate, max fleet size & unsatisfied demand | Bus-rail | Real world | Real world | Routing & service frequency | GA |
| Shrivastava and O'Mahony (2009) | Min. sum of transfer cost, travel cost and operation cost | Max/min load rate, max fleet size & unsatisfied demand | Bus-rail | Real world | Real world | Routing & service frequency | GA-a heuristic |
| Lin and Wong (2014) | Max. service coverage, minimax travel time, min route length | Route connectivity, subtour prevention, max. travel time of a route & relationships between route layout and service coverage | Bus-metro | Real world – grid network | Fuzzy numbers | Routing & trip assignment | LINGO |
| Pan et al. (2015) | Max. the number of served passengers & min. operational cost | Fleet size, flow, subtour, vehicle capacity, max travel time, min route length & demand point-vehicle assignment | Bus | Real world | Given | Routing & service area | A gravity based heuristic algorithm |
| Almasi et al. (2015) | Min. sum of waiting, in-vehicle and access costs, operation cost & social cost | Route feasibility, route-station matching, min/max route length, min/max service frequency, available fleet size & users satisfaction | Bus-Rail | Non-idealized test network | Given parameter | Routing & service frequency | GA, PSO & imperialist competitive |
| Gschwender et al. (2016) | Min. sum of waiting, in-vehicle and operation costs | Vehicle capacity | Bus | Stylized public transport network | Given parameter | Fleet size & vehicle size | - |
| Zhu et al. (2017) | Max, potential users demand | Travel time, number of links & link closure | Bus-Rail | Real world grid network | Demand estimation function | Routing & user flow distribution | GA |
| Li et al. (2018) | Min. sum of users access and in-vehicle costs, and distance traveled by bus | Number of pick-up locations, user-location assignment, vehicle-pick-up location matching, subtour, vehicle capacity, unserved demand, travel time, route length, user-line assignment & stop-station matching | Bus-rail | Real world | Real time data | Routing & pickup locations | Out-of-the-shelf solver & GA |
| Sun et al. (2018) | Min. operation cost & max. users satisfaction | Demand location-vehicle matching, stop-station matching, max/min travel time, subtour & vehicle capacity | Bus-Rail | Real world | Given parameter | Vehicle assignment | Bat algorithm, CPLEX & group search optimizer |
| Huang et al. (2018) | Min. sum of in-vehicle and waiting costs & min. sum of waiting and operation costs | Available fleet | Bus-Rail | A demand distribution function | Test network | Bus lines, Hubs, service frequency | Artificial bee colony algorithm |
| Sun et al. (2019) | Min. total travel distance | Demand point-vehicle matching, flow, stop-station matching, preferred pick up time period, max travel time and travel distance, & vehicle capacity | Bus-Rail | Fuzzy demand | Real world | Routing and arrival timing | ACO, PSO & GA |
| Park et al. (2019) | Max. cosine similarity | Number of intermediate stops | Bus | Real world | Real GPS data | Routing & stop location | Couene algorithm |
| Yang et al. (2020) | Min. access, waiting, in-vehicle. Operation, line operation, fleet & station costs | Vehicle capacity | General, Bus-Rail | Test network | Spatially heterogeneous demand | Trunk service headway, feeder service headway, station location & line spacing | GA |
| Wei et al. (2020) | Min. sum of walking, in-vehicle & waiting costs | Maximum walking distance, upper/lower bound for service frequency & load factor | Bus | Real world | Real mobile phone data | Routing & frequency setting | GA-Dijkstra search |
| Wang et al. (2020c) | Min. sum of operation cost and users travel time | Vehicle capacity, users' time window & bus running time | Bus | Real world | Reserved and real-time demands | Routing & service scheduling | GA |
| Galarza Montenegro et al. (2021) | Min. sum of total vehicles travel time, access time & difference between actual arrival time and desired travel time | - | Bus | Test network | Given parameter | Stops to visit, user-vehicle assignment, users arrival time & bus departure time | Large neighborhood search algorithm |
| Li et al. (2021) | Min. average users' delay time | Vehicle capacity, time continuity, time window for service, working time, & users arrival and boarding time | Bus-train | Real world | Given parameter | Routing & passenger assignment | ALNS |
| Wei et al. (2021) | Min. in-vehicle cost, Min. walking cost | Maximum number of bus stops, demand point-stop assignment & max walking distance | Bus | Real world | Cell phone data | Routing, bus location. Users route choice | Ant colony algorithm |

### 2.3. Service pricing and fare management

Service pricing and fare management is one of the major subcategories where researchers have applied mathematical modeling and optimization algorithms to study the related problems. For example, Daskin et al. (1985) developed a mathematical model to design a fare obtaining system considering both the agency's and users' benefits. They assumed that the service fare must be a summation of a fixed price, a traveled distance based price, and charges for the transfers. The objective of the proposed model was to maximize the agency's profit where the constraints were related to the ridership level and fare structure. Considering a single corridor bus transit service with different time-dependent service patterns and different vehicle capacities, Delle Site and Filippi (1998) proposed an optimization model to find optimal features of this transit system. The objective was to maximize the difference between users' benefits and the uncovered agency's costs. Tsai et al. (2008) developed a mathematical model to investigate the joint frequency and service fare setting problem to find optimal values for the decision variables leading to the maximum agency profit. They assumed that fare depends on travel distance, and demand depends on fare and waiting time. They considered fleet size and service level as the constraints. To solve the proposed model, an algorithm was suggested and applied to a real transit system in Taiwan. Li et al. (2009) considered two different market structures for a transit network, including a monopoly and an oligopoly, and developed bi-level programming to study the fare management problem. At the upper level, the agency was the decision maker aiming to obtain the optimal fare leading to its maximum profit, while the lower level model was related to the route choice problem of users. They claimed that uncertainty in the network will be result of uncertainty in line-haul, in-vehicle travel time or dwell times at stops. They considered the uncertainty in the network by capturing them in the disutility functions.

To obtain the optimal value of operational characteristics of a PBTS, including service frequency, bus sizes, stop spacing, the method of fare payment, and running speed, Tirachini and Hensher (2011) proposed an optimization model with the objective of minimizing the summation of users' and agency's costs. To this end, they considered a single corridor bus transit service, and showed that, in most cases, off-board fare payment is the best payment option, and cruising speed would be increased proportional to the logarithm of demand. Based on the users' choice model, Borndörfer et al. (2012) developed a nonlinear optimization model to evaluate the different fare systems, considering different objective functions and operational strategies. The model was applied to the transit network of Potsdam, Germany, to find the optimal fare system for this city.

Zhang et al. (2014) proposed a bi-level programming model to study service fare setting problems for various types of public transit services, taking uncertainty in supply and demand into account. The optimal fare level was obtained using the upper level model with the objective of minimizing the total agency's cost and the traffic assignment problem was solved at the lower level. To deal with the complexity of the proposed model, a heuristic algorithm was suggested. The results indicated that weather conditions and time values will significantly affect the optimal service fares. In particular, they claimed that adverse weather conditions will affect the uncertainty of the demand for public transit services. In addition, as a result of adverse weather conditions, the mode choice behavior of the users and the results of the traffic assignment will be changed. Tang et al. (2017) studied the problem of fare structure setting and operations planning for a public transit service where the objective was to maximize the total social welfare. They considered different operational strategies such as short turns, full routes and limited stops, and mixed services to serve existing demand in the network. In addition, the impacts of the crowding level inside the vehicle and service fare on demand were taken into account in the modeling process. They applied the proposed model to a real network in Dalian, China. Developing a bi-level programming, Correa et al. (2017) investigated different fare inspection strategies in public transit services. The objective of the upper level model was to identify the probabilities of users' inspection at the different locations and in the second level model, the fare-evading users optimized their route choice knowing the travel time on each route and inspection probabilities in each location. They considered two different strategies for users' routing. For the first strategy, they assumed users would stay on the selected route, but in the second strategy, users can update their route in any step of the travel based on their perceived information. In addition, they considered two different fare structures, including fixed and flexible fares.

Taking an elastic demand for a public transit service into account, Chowdhury and Chien (2019) developed an optimization model for joint optimization of fare and service frequency. The objective of the proposed model was to maximize the agency's profit considering different constraints related to the service level. They also considered the possibility of giving a discount on the service fare in order to encourage users to use public transit service. Finally, the proposed model was applied to solve some numerical examples, and the results were discussed. Kamel et al. (2020) focused on the problem of the time dependent fare structures for large scale public transit networks and developed an optimization model to investigate this problem. They considered the impacts of fare level on the users' mode and route choice, and their departure time in the modeling process and applied the proposed model to the case of the Greater Toronto Area, Canada. They found that time varying fares can help agencies divide the demand during peak hours. In another similar study, Guo et al. (2021) developed a model to study the time dependent fare setting problem for a transit network with an elastic time-dependent location-dependent demand pattern. To this end, they considered a transit line providing many-to-many transit services and optimized the service fares, frequency, and vehicle and fleet sizes. Their results indicated that a time-dependent fare system would prevent cross-subsidization during different times.

Table 3 represents the main features of the reviewed papers in this subsection. According to this table, maximization of the agency's profit and maximization of social welfare are most commonly considered objective functions in this section. Service fare and fare collection system are the main decision variables in this section, while the main constraints are related to the vehicle capacity, budget, service frequency, and service level.

**Table 3.** Main features of the reviewed papers in subcategory 3.

| Reference | Objective Function | Constraints | Transit mode | Network structure | Demand Pattern | Decision Variables | Solution Approach |
|---|---|---|---|---|---|---|---|
| Daskin et al. (1985) | Max. estimated revenue | Min ridership, attributes of the fare structure | Bus | Real world | Function | Transit fare | - |
| Delle Site and Filippi (1998) | Max. users benefits by lower average waiting time minus operators costs not covered by fares | Min frequency, capacity | Bus | A bus corridor | Given O-D matrix | Turn back positions, vehicle size, service frequency, relative offsets of full length and short turn line, fare | - |
| Tsai et al. (2008) | Max. total profit | Service capacity, fleet size | General | Real world | Elastic demand | Distance based fare, service frequency | |
| Li et al. (2009) | Max. agency's profit, max. social welfare | - | Bus, general | A small example network | Elastic demand | Fare structure, users mode choice behavior | Method of successive average, SA, sensitivity analysis based algorithm |
| Tirachini and Hensher (2011) | Min. sum of users' access and waiting, in-vehicle costs, agency's land, infrastructure, station, personnel and running cots | Max/min capacity, Max/min frequency, Max/min running speed | Bus-BRT | A bus corridor | Given parameter | Fare collection system, service frequency, capacity, running speed, stop spacing, modal split | Codded in MATLAB |
| Borndörfer et al. (2012) | Max. revenue, Max. profit, Max. demand, Max. users benefit, | Service level, subsidies, service frequency, | General | Real world | Function | Fare | GAMS |
| | Max. social welfare | budget, capacity | | | | | |
| _ENREF_80Zhang et al. (2014) | Min. the transit network's total travel and operation cost | Max/min flat fares | Multi-modal: bus-train | An example network | Elastic demand | Fare structure, users mode choice | A heuristic, Method of Successive Averages |
| Tang et al. (2017) | Max. social welfare | Vehicle capacity, number of consecutive skipped stops, potential demand | Bus | Real world | Elastic demand | Fare structure, operational strategies | GAMS: DICOPT |
| Correa et al. (2017) | Min. expected cost | - | General | Real world | Random uniform demand | Users inspection probabilities, route choice behavior, fare structure | Exact, an approximation search algorithm |
| Chowdhury and Chien (2019) | Max. total profit (waiting time, in-vehicle time, transfer time, operation costs) | Service constraints | Bus | A test network | Elastic demand | Fare, headway | Powell's algorithm |
| Kamel et al. (2020) | Min. average pf total travel time (in-vehicle, access, egress, waiting and walking times) | - | Multi-modal | Real world | Real data | Time dependent fare, mode choice behavior, departure time | GA |
| Guo et al. (2021) | Max. social welfare | Fleet size, vehicle capacity | General | A transit line | Elastic and spatially distributed | Fare, headway, vehicle capacity, fleet size | - |

## 2.4. Fleet management

Another identified subcategory includes published studies focusing on fleet management problems. These problems can be related to fleet sizes, vehicle sizes or vehicle allocation. Han and Wilson (1982) investigated the bus allocation to the different transit routes using a mathematical programming with the objective of minimizing users' waiting time and in-vehicle crowding. They considered the route choice behavior of the users in the modeling process. The main constraints were related to the available fleet size and the needed capacity for each route. An innovative algorithm was developed to solve the model, and it was applied to the real case of the Cairo bus system. Considering a transit corridor and different operation periods, Delle Site and Filippi (1998) developed an optimization model to obtain the optimal turnback positions, size of the buses, headways, service fares, regular arrival locations, and full-length and short-turn line relative offset, in each period. Maximizing the difference between users' benefits and the agency's costs that are not covered by service fares was the objective function of the model. Chakroborty et al. (2001) developed a GA to solve the vehicle allocation and scheduling problems for a transit service. They used GA to find optimal or near to optimal solutions for the understudy problem.

Considering a bus feeder shuttle transit service that serves passengers communicating between a recreation center and a major public transit station, I-Jy Chien (2005) used mathematical models to obtain service frequency, bus size, and optimal route choices. Service capacity, available budget, and vehicle availabilities were the main constraints of the model. They suggested an approach to solve he model and applied it to a real case. Chowdhury and Chien (2011) developed a mathematical programming with the objective of minimizing the summation of users' and agency's costs in a public bus transit service, including a major hub and multiple routes. They optimized the vehicle capacity, service frequency and slack time subject to available capacity and coordination levels. Ibeas and Ruisanchez (2012) proposed a bi-level programming to calculate bus size and headways for each of the bus transit routes, considering vehicle capacity as the constraints and observed demand. The objective function of the upper level was to minimize total users' and agency's costs, while the lower level model was a public tranist assignment model. In another similar study, Ruisanchez et al. (2012) jointly optimized service frequency and bus sizes for each transit route using a bi-level model with a similar structure to the previous study. They applied Hooke-Jeeves or Tabu search algorithms to solve the proposed model for the real case of Santander, Spain, and compared the performance of these algorithms.

Considering the summation of waiting and crowding costs of the users and the size agency's costs, including costs of vehicle capacity, unused capacity, and lost sales, Herbon and Hadas (2015) used a newsvendor model to obtain vehicle capacity and service frequency for a transit service. They considered an upper bound for the waiting time and an upper bound for the vehicle capacity as the main constraints. For the understudy network, a stochastic demand was assumed, and an algorithm was proposed to solve the model in a reasonable evacuation time. They applied the proposed approaches to a real case study and conducted sensitivity analysis to find insightful results. Taking minimizing users' waiting time and minimizing agency's operations costs as two objective functions, Gkiotsalitis et al. (2019b) developed a multi-objective programming to study the bus allocation problem. They considered two lining strategies, including short-turning and interlining lines. Using exterior point penalties, the proposed model was converted to an unconstrained model, and to deal with the complexity of the proposed model, a GA was suggested to solve it in a reasonable running time. Suman and Bolia (2019) used optimization models to reassign the existing buses to the transit routes, aiming to minimize the required fleet size to serve current and predicted demand in the network. The main objective of this study was to deal with the in-vehicle overcrowding problem by assigning current service level and providing optimal operation strategies for the future.

Liu et al. (2019) focused on the feeder bus transit system that feeds a metro station. It was assumed that users could also use bike-sharing services to communicate between their origin and metro station and vice versa, and service frequency during the morning follows a fixed pattern while for the evening it follows a dynamic demand responsive pattern. They proposed a multi-objective programming to obtain fleet size and service frequency for this system. The objectives were minimizing users' waiting time and maximizing agency's profit, and the constraints were related to bus capacity and users' mode choice behavior. They proposed two metaheuristics to solve the proposed model, including NSGA-II and MPSO, and then applied them to the transit network of Chengdu, China. Their results showed that MPSO performed better than NSGA-II, from different perspectives. Jara-Díaz et al. (2020) investigated the possibility of using multiple a fleet strategy to serve passengers in a PBTS, where a regular fleet serves passengers during the whole operation time and another fleet supports the system by serving during only peak hours. The vehicle sizes in each fleet could be different. Their results indicated that joint optimization of the operation of both fleets would yield better results compared to independently optimization. Also, they showed that a multiple fleet strategy is more efficient in cost than a single fleet. Taking a time dependent demand pattern into account, Zhang et al. (2021b) developed a two-stage heuristic algorithm to study the optimal schedules for vehicle utilization, vehicle purchases, and vehicle retirements. Applying the proposed algorithm to a set of numerical examples, they made a comparison between the performance of their algorithm and other existing algorithms in the literature.

Table 4 represents the main features of the reviewed papers in this subsection. According to this table, waiting time, crowding, in-vehicle time, transfer time, access costs, operation costs, and unused capacity are the main cost factors considered by different studies in this section. While constraints are mostly related to the available fleet size, service frequency, budget, service level, and demand, the most commonly considered decision variables are vehicle size, service frequency, and vehicle allocation. Looking at the solution approaches, it can be seen that heuristics and metaheuristics are the main approaches to solving the proposed models in this section. In particular, GA is the most popular one. Although conducted studies have tried to capture real world conditions by taking realistic transit network structures and demand patterns into account, none of them has considered stochasticity in the modeling process. In the real world, we always face an uncertainty problem in transit networks. The uncertainty can be considered for travel time, demand, users' mode choice, etc. Addressing the uncertainty problem in the proposed models can lead to more realistic results.

**Table 4.** Main features of the reviewed papers in subcategory 4.

| Reference | Objective Function | Constraints | Transit mode | Network structure | Demand Pattern | Decision Variables | Solution Approach |
|---|---|---|---|---|---|---|---|
| Han and Wilson (1982) | Min. waiting time and crowding at max load point | Fleet size, route capacity | Bus | Real world | Given O-D matrix | Vehicle allocation | Heuristic |
| Delle Site and Filippi (1998) | Max. users benefits by lower average waiting time minus operators costs not covered by fares | Min frequency, capacity | Bus | A bus corridor | Given O-D matrix | Turn back positions, vehicle size, service frequency, relative offsets of full length and short turn line, fare | - |
| Chakroborty et al. (2001) | Min. sum of total waiting time and total transfer time | Resource, min/max stopping time, policy headway, max transfer time | Bus | Example network | Given parameter | Fleet size, service frequency | GA |
| I-Jy Chien (2005) | Min. sum of access, wait, in vehicle & operational cost | Vehicle schedule, bus availability, service capacity, budget | Bus | Real world | Uniform distribution | Fleet size, service frequency, route choice | - |
| Chowdhury and Chien (2011) | Min. sum of waiting, transfer, in-vehicle, operational costs | Capacity | Bus | Real world | Uniformly distributed many-to-one/one-to-many demand pattern | Vehicle size, service frequency, slack time | Powell's algorithm, Gaussian quadrature |
| Ibeas and Ruisanchez (2012) | Min. sum of personnel cost, fix operational cost, access time, in-vehicle time, transfer time, waiting time | Bus capacity | Bus | Real world | Varying demand obtained by simulation | Vehicle size, service frequency, public traffic assignment | Hooke-Jeeves algorithm, ESTRAUS |
| Ruisanchez et al. (2012) | Min. sum of operational costs, waiting time, access, in-vehicle & transfer costs | Capacity | Bus | Real world | Fix total demand but varying along the routes | Service frequency, bus size | Hooke–Jeeves, Tabu search algorithms |
| Herbon and Hadas (2015) | Min. waiting, overcrowding, vehicle size, empty seats & lost sales costs | Max average waiting time, max vehicle capacity, service level | Bus | Real world | Real data | Service frequency, vehicle capacity | Lagrangian based search algorithm |
| Gkiotsalitis et al. (2019b) | Min. users waiting time, min. operational cost | Service level, fleet size, waiting time | Bus | Real world | Real data | Lining, service frequency, fleet allocation | GA |
| | | maximum threshold | | | | | |
| Suman and Bolia (2019) | Max. passenger-kilometers, Min. sum of capital and operation costs | Demand, resource, policy headway, capacity | Bus | Real world | Given parameter | Fleet size, vehicle allocation, current service level | CPLEX |
| Liu et al. (2019) | Min. average users waiting time | Fleet size, trip time, user flow, vehicle flow, service capacity, operating time | Bus-bike-metro | Real world | Endogenous demand | Fleet size, service frequency | NSGA-II, MPSO |
| Jara-Díaz et al. (2020) | Min. sum of capital cost, operation cost, waiting and in-vehicle costs | Capacity constraints | Bus | Single line | Different off-peak and peak demand | Service frequency, single or mixed fleet strategies, bus capacity | - |
| Zhang et al. (2021b) | Min. sum of total discounted vehicle purchase cost, the total discounted O&M cost, the total discounted | Demand, cumulative mileage, vehicle types | General | Example network | Time varying demand | Vehicle utilization schedule, vehicle purchase and retirement schedule | Two heuristics |

### 2.5. Dispatching policy

Dispatching policy setting is one of the most important and, at the same time, most complicated problems in PBTNDPs&OP using any methodology. This problem is the main focus of a large portion of the identified papers. This subject is also one of the earliest subjects investigated by researchers using mathematical programming and optimization algorithms. However, since there are a number of already published literature review papers that have focused on this problem, in this section only the selected papers published after 2016 will be reviewed.

Wu et al. (2016a) studied the robust schedule coordination planning aiming to manage the disruptions in the transit service using semi-flexible holding strategies. To this end, they integrated an analytical model and a stochastic mixed-integer programming model considering three operation scenarios, including uncoordinated, departure punctual, and departure delayed controls. A B&B algorithm was also suggested to cope with the complexity of the proposed model. It was applied it to a group of numerical instances to simultaneously optimize service headway and slack times. Wu et al. (2016b) provided multi-objective programming to study the re-synchronizing of public bus timetable problems with the objective of striking a balance between the total number of benefited users by smooth transfers and the maximum difference between the designed schedule and the real schedule. They considered the irregularity of the service headway, the variability of the travel demand due to headway irregularity, flexible coordination, and existing dispatching policy in the modeling process. Proofing that the proposed model is an NP-hard model, they suggested an NSGA-II to handle the complexity of the model for the real size networks. Comparing with the full enumeration methodology, they showed that the proposed NSGA-II would provide better Pareto solutions at an acceptable running time. Fouilhoux et al. (2016) developed four modifications of an existing mixed-integer optimization model to solve the coordinated timetabling problem for a bus transit service. They tried to reduce the users' transfer costs and bus bunching at the stops. By solving numerical examples, they analyzed different perspectives of proposed models, including solution quality and execution time. In order to design a bus dispatching policy leading to an acceptable level of waiting time for the users without any significant increases in the operation costs, Canca et al. (2016) proposed mathematical modeling to evaluate different short-turning strategies. The objective of the proposed model was to minimize users' waiting time by obtaining the optimal turn-back locations and service offset with respect to the considered service level for the transit system. They applied the proposed methodology to a real case study and discussed the results.

Taking the route choice behavior of the users into account, Liu and Ceder (2017) investigated the coordinated timetabling problem using a bi-level optimization model. They suggested a sequential search method based on an introduced deficit function in order to find Pareto solutions for the proposed multi-objective integer model. They integrated this algorithm with network flow and shifting bus departure time methods to solve a set of numerical instances. Asgharzadeh and Shafahi (2017) developed an optimization model to obtain the optimal real-time holding time of buses that leads to the minimum onboard and at-station waiting costs of the users. To handle the complexity of the proposed model that has a root in the real-time decision making process, they suggested a gradient descent based heuristic algorithm and applied it to the BRT network of Mashhad, Iran. The results showed that, under optimal values obtained by the proposed model, the waiting time of the users would be decreased by up to 8.65%. Considering a multimodal transit network and a Normal distribution function for bus travel times, Wei and Sun (2017) provided a bi-level optimization model to address the optimal bus timetabling problem. The timetable coordination problem was solved in the lower-level model, while the upper-level model was responsible for the trip assignment problem with the objective of minimizing operation costs. Finally, they used an Estimation of Distribution Algorithm to solve the proposed bi-level programming in a reasonable computation time. In another similar study, Gkiotsalitis and Maslekar (2018) studied the PBTS timetabling problem using constrained and unconstrained optimization models, aiming to minimize the users' waiting time fluctuations. They considered different operational facts in the modeling that were ignored in the literature, such as rest time, mealtime, etc. An evolutionary algorithm based on the exterior point penalty scheme was used to solve the proposed models and applied to a large real PBTS in a city in the Asia Pacific area.

Petit et al. (2018) analyzed the benefits of having a standby bus fleet in the PBTS to replace an early or late bus, aiming to maintain schedule adherence. They addressed this problem with a dynamic optimization model where the objective was to obtain optimal dispatching policies leading to the minimum total users' and agency's cots. They demonstrated, through a number of numerical examples, that having a standby fleet will significantly assist the agency in managing the irregularity issue in service frequencies. It also was shown that compared to the slack-based models, the proposed model obtains more efficient results. In another similar study, Petit et al. (2019) discussed the advantages of having a standby bus fleet in a multiple line PBTS using stochastic dynamic optimization modeling. Considering time-independent or time-varying assumptions, the proposed model was used to find the optimal dispatching policies and optimal decisions about the retired buses that minimize the total users' and agency costs. Gkiotsalitis and Alesiani (2019) focused on designing a robust coordinated bus timetable considering different types of uncertainties in the network. They considered fluctuations in both travel demand and travel time in the modeling process and provided an optimization model with the objective of obtaining an optimal timetable leading to the minimum possible loss in worst-case scenarios. To cope with the computational complexity of the proposed model, a sequential quadratic technique based GA was suggested and was validated by applying it to a bus transit line in Singapore. In order to have efficient timetable coordination in a PBTS, Gkiotsalitis and Cats (2019) proposed a periodic methodology to simultaneously find the optimal vehicle holding times for all running vehicles in each optimization time slot. To this purpose, they provided nonlinear discrete mathematical programming with respect to the undesirable impacts of holding strategies on the in-vehicle travel costs. Because of the high complexity of the bus holding problem, they also suggested an alternative optimization method and applied it to the real case of Stockholm, Sweden.

Li et al. (2019b) considered both demand and travel congestion fluctuations and developed robust dynamic programming based on the state-feedback scheme to study the bus holding problem. Their objective was to minimize bus bunching in the network by decreasing the irregularity in the service frequency and increasing schedule adherence. The bus's motion was also investigated with a state-space model. To deal with computational complexities, the proposed non-linear programming was reformulated as a convex model that was able to solve the bus holding problem for real size networks in reasonable running time. The bus scheduling problem for a transit service with multiple vehicle types considering both passengers and bus flow has been studied by Li et al. (2019a). They developed an integer linear model for this problem by taking range and refueling constraints, and emission costs into account, aiming to find the optimal service frequency and refueling facility locations leading to minimum total users' and agency costs. A simplified version of the proposed model was used to find approximate solutions for real-sized networks and then was applied to the bus network of Hong Kong.

Gkiotsalitis (2019) tried to capture the impacts of the hourly changes in travel demand and other factors in the transit network in the bus service scheduling problem. To this end, they proposed a model to periodically obtain the optimal service schedules for both running buses and remaining trips. They considered layover times and capacity limits as the main constraints of the proposed model. To find near to optimal solutions for the proposed model in an acceptable execution time, they also suggested a sequential hill climbing based algorithm and tested it on a bus line in Denver, USA. Chen et al. (2019) proposed mixed-integer linear programming to simultaneously optimize the service frequency and bus size. They also suggested an algorithm based on dynamic programming to solve the proposed model. To handle the complexity issue, they added some valid inequalities to the algorithm, which were related to the queue length and bus size. By this, they limited the search space for the algorithm and, as a result, decreased the computation time significantly. Solving some numerical instances, it was shown that the proposed algorithm would outperform other existing algorithms in the literature; in addition, they showed that inputs would impact the dynamic programming results in both oversaturated and unsaturated traffic conditions.

Rinaldi et al. (2020) considered a transit network that serves users with a mixed e-bus and hybrid bus fleet, and then investigated the service frequency setting problem using a mixed-integer linear optimization model. They analyzed the benefits of integrating e-buses with existing regular or hybrid PBTSs from the operational cost perspective. The proposed model was used to discuss the mentioned problem for the public bus transit network of the city of Luxembourg and found that gradually moving from regular bus transit services to e-bus based transit systems will benefit the agency with reduced operational costs. However, when the e-buses out-numbered the regular buses, the marginal savings would decrease. The robust bus service scheduling problem considering fluctuating traffic congestion has been studied by Wang et al. (2020a). They proposed a three-phase optimization procedure for this problem, where first a set of bus blocks was produced, second, an NSGA-II was applied to form a set of Pareto solutions, and finally, a dynamic dispatching method was used to evaluate and rank the non-dominated solutions. They applied the model to a real case study and compared the obtained results with the results of the other existing models in the literature.

Aiming to minimize the difference between actual headway and designed headway, Bie et al. (2020) proposed a dynamic mathematical programming to study the real-time service headway setting problem for a single high-frequency bus route. Vehicle speed adjustment and intersection signal resetting are two main strategies in their bus holding model. They also suggested a formula for bus travel time estimation and a modified GA to solve the proposed methodology for a real case in Meihekou City, China. Gkiotsalitis and Van Berkum (2020) developed a headway setting scheme for each rolling horizon in a PBTS. To that end, they used non-linear optimization programming to model the problem, which was shown to be a convex model capable of finding the optimal solution(s) in a reasonable amount of time. Applying the methodology to a bus line in Denver, USA, the performance of the proposed model was proved to be better than the performance of the existing myopic models. Gkiotsalitis (2021) proposed a dynamic nonlinear integer optimization model to obtain the optimal service reschedule and bus holding strategies for a PBTS, leading to the minimum excessive waiting time of the users. They showed the proposed model is an NP-hard model and provided a heuristic algorithm to deal with this complexity. Solving numerical examples, the performance of the suggested algorithm was compared to the performance of the SA, B&B, and hill-climbing algorithms. The model and algorithm were also applied to a major PBTS in the Asia Pacific as a real case study.

Focusing on the service frequency setting problem considering congestion level in the common transit lines in PBTS, Tian et al. (2021) developed a triple-level mathematical programming taking both transit route choice behavior and congested common transit line choice behavior of the users into account. In the upper-level model, the optimal service frequency was obtained with the objective of minimizing the total agency's and users' costs. The second level model was related to the route choice problem of the users, in which the objective was to minimize their travel costs. Finally, the lower level model was responsible for obtaining the equilibrium scenario on the congested common transit lines. The triple-level model is then converted to a single-level model. Two strategies were suggested to handle the complexity of the final model: first, the model was reformulated as a mixed-integer linear optimization model capable of finding exact solutions, and second, using a surrogate optimization algorithm. Silva-Soto and Ibarra-Rojas (2021) optimized the service frequency and timetabling with coordination between different bus lines at the common stops by developing a mixed-integer linear optimization model. The objective was to minimize the average waiting time of the users at the common stops and total operating costs. To estimate the average waiting time of the users, they considered time-dependent variables. They used CPLEX, SAW, BRKGA, and BRKGA-NSGA-II algorithms to solve different versions of the proposed model for a group of examples.

Liang et al. (2021a) developed a multi-objective optimization programming for the service frequency setting problem for a limited-stop bus transit service. The objectives were minimizing the users' costs and minimizing the agency's costs. They suggested a prioritization technique to evaluate the unsatisfied demand and an evolutionary algorithm to solve the model. The proposed methodology was used to solve the problem in a numerical example and a real case. As a result, they found that the frequency and number of skipped stops would have different impacts on the fleet size, transfer costs, waiting costs, and in-vehicle costs. Considering four different operation strategies for a single bus transit network that was serving a fluctuating travel demand, Tang et al. (2021) proposed a model for the timetabling problem. The strategies were full route operation, limited-stop, short-turn, and mixed operations. The objective of the proposed model was to minimize the total cost in the network by modifying the departure time of the buses. The total cost function was obtained as the summation of the users' costs caused by the difference between modified and planned schedules obtained in the full route operation strategy. The proposed method was validated by applying it to a test network and the real data of Dalian, China. The results showed that the combined strategy could decrease the total cost in the network by up to 50%. In an integrated rail-bus transit network, the problems of bus bridging, timetabling and service scheduling were formulated by Wang et al. (2021). The main focus of the proposed bi-level model was on the transfer costs and users' tolerance toward connection maintenance. The upper-level model was related to the timetabling problem with the objective of the number of failed transfers and waiting time minimization, while the bus scheduling problem was formulated at the lower level, taking operation cost minimization as the objective function. A SA algorithm was also proposed to handle the computational complexities for the real size networks, which was then applied to solve the problem for the real case of Shanghai, China.

Table 5 represents the main features of the reviewed papers in this subsection. Based on this table, most of the recently published papers on dispatching policy have considered waiting costs, transfer costs and failures, delay costs, schedule adherence, headway irregularity, synchronization, and operation costs as the main factors in their model. Timetable, service frequency, holding strategies, and dwell time are the main decision variables in this section. Also, min/max headway, min/max holding time, departure time, and vehicle capacity are the most commonly considered constraints. A large variety of optimization algorithms, heuristics, and metaheuristics have been applied to find near to optimal solutions for the proposed models in this section. In particular, B&B, GA, and SA have been used in various studies. Despite the large number of published studies in this subcategory, there are not enough attempts to address dispatching policies and headway setting problems for the PBTSs considering the introduction of emerging technologies. For example, it is predictable that the operation of integrated E-bus and traditional bus based transit networks and E-bus based transit networks will be different from the operation of conventional PBTSs. Therefore, future studies can focus on these problems.

**Table 5.** Main features of the reviewed papers in subcategory 5.

| Reference | Objective Function | Constraints | Transit mode | Network structure | Demand Pattern | Decision Variables | Solution Approach |
|---|---|---|---|---|---|---|---|
| Wu et al. (2016a) | Min. sum of operation cost, waiting cost, transfer cost, delayed connection cost, induced trading costs, monetary incentive for schedule recovery effort & extra waiting costs | Max/min headway | Bus | Test network | Given parameter | Planning headway, slack time, flexible departure control strategy | B&B |
| Wu et al. (2016b) | Minimax difference between existing departure time and new departure time, max. number of benefited users by smooth transfer | Departure time and planning period, max/min headway, min/max separation time between synchronized bus arrivals | Bus | Test network, real world | Headway-sensitive demand | Timetable | NSGA-II, enumeration method |
| Fouilhoux et al. (2016) | Max. total number of synchronization | Min/max headway, departure time window, min/max difference | Bus | Test network | Rate for each node | Timetable | CPLEX : four valid inequalities |
| | | between arrival times | | | | | |
| Liu and Ceder (2017) | Min. sum of users in-vehicle, waiting time, transfer time, load discrepancy cost, min. fleet size | Fleet size, departure time | General, bus | Example network | O-D demand matrix | Dispatching policy: timetable synchronization | A deficit function (DF)-based sequential search |
| Asgharzadeh and Shafahi (2017) | Min. sum of on-board and on-stop users' waiting time | Number of on-board users, capacity, upper bound for holding time | BRT, bus | Real world | Real time | Dispatching policy: optimal holding time | A gradient descent algorithms |
| Wei and Sun (2017) | Min. operational costs | Parking capacity | Bus | Real world | Real data/GPS | Timetable | Estimation of distribution algorithm |
| Gkiotsalitis and Maslekar (2018) | Min. the waiting time variations of users at stops | The required mealtime and resting time breaks, the required headway ranges between successive bus trips | Bus | Real world | Temporal variations | Timetable | An exterior point penalty scheme |
| Petit et al. (2018) | Min. sum of operational costs, substitution cost, waiting time, delay costs | Substitution time, maximum number of standby assignment in each time period | Bus | Example network | Time varying | Dispatching policy | Approximate dynamic programming algorithm |
| Petit et al. (2019) | Min. sum of operational costs, substitution cost, waiting time, delay costs | Substitution time, maximum number of standby assignment in each time period | Bus | Real world | Uniform (density factor) | Dispatching policy | Approximate dynamic programming algorithm |
| Gkiotsalitis and Alesiani (2019) | Min. service irregularity | Layover time, first and last trip departure time, max headway | Bus | Real world | Uncertain demand | Timetable | GA |
| Gkiotsalitis and Cats (2019) | Min. the variance of the actual users' waiting times from the planned waiting times | Slack time, travel time limits for on-board users, upper limits of bus holdings | Bus | Real world | Time-dependent location dependent demand | Dispatching policy: bus holding times | B8B, an alternating minimization approach |
| Li et al. (2019a) | Min. sum of operational costs, emissions costs, | Range, refueling, bus | Bus | Real world | Given | Dispatching policy | CPLEX |
| | lost demand penalty, waiting costs, in-vehicle costs, access costs | connectivity, capacity | | | | | |
| Li et al. (2019b) | Schedule adherence and headway regularity | State constraints | Bus | Real world | Uncertain time varying demand | Dispatching policy: bus holding strategies, speed adjustment | Adaptive control |
| Gkiotsalitis (2019) | Min. excessive waiting times | Original timetable, dispatching time interval, layover time, first and last trip departures | Bus | Real world | Time-dependent | Service frequency, bus holding strategy | Heuristic, GA, SA, hill climbing, B&B |
| Chen et al. (2019) | Min. the total waiting cost and energy cost | Vehicle capacity, Passenger flow conservation, safety headway | General | Real world | Time-dependent parameter | Dispatching policy, vehicle capacity | DP algorithm |
| Rinaldi et al. (2020) | Min. total operational cost | Max trip at a time, trip-bus matching, trip departure time, energy level, charger-e-bus matching, discharging rate | E-Bus & hybrid bus | Real world | - | Dispatching policy | CPLEX |
| Wang et al. (2020a) | Min. number of uncovered start times in the timetable, Min. number of drivers | - | Bus | Real world | - | Dispatching policy | GA |
| Bie et al. (2020) | Min. sum of difference between actual headway and dispatching headway, and ratio of signal cycle length scaling | Speed, signal cycle scaling | Bus | Real world: single line | - | Dynamic headway, speed and signal adjustment | GA |
| Gkiotsalitis and Van Berkum (2020) | Min. headway variance around the target headway | Schedule sliding, trip-bus matching | Bus | Real world | Given parameter for each stop | Dispatching policy, dwell time, bus arrival time, busload, stranded users | Exact |
| Gkiotsalitis (2021) | Max. synchronization (max. number of simultaneous bus arrivals at the connection nodes) | Min/max headway, first/last departure time | Bus | Example network | - | Dispatching policy, users transit route choice, transit service line choice | Heuristic |
| Tian et al. (2021) | Min. sum of users waiting and in-vehicle costs & operational costs | Max/min frequency | Bus | Example network | Given O-D demand matrix | Dispatching policy | Exact, a surrogate optimization approach |
| Silva-Soto and Ibarra-Rojas (2021) | Min. the average waiting time at synchronization points, Min. operational costs | Departure time window, consecutive arrivals | Bus | Example network | Arrival rate for each stop | Timetabling & service frequency | CPLEX, BRKGA, BRKGA-NSGA |
| Liang et al. (2021a) | Min. fleet size, min. transfer, in-vehicle and waiting costs | Max/min service frequency, vehicle capacity | Bus | Real world | Given O-D demand matrix | Service frequency, service pattern | A cooperative co-evolutionary multi-objective evolutionary algorithm |
| Tang et al. (2021) | Min. change in the users' travel time caused by vehicle departure time adjustment | Shortest path, connectivity | Bus | Real world: single line | Temporally and spatially distributed demand | Dispatching policy: operation strategy | A shortest-path algorithm |
| Wang et al. (2021) | Min. sum of users waiting time & the number of transfer failures, Min. bus operation cost | Max/min headway, departure time, transfer connection | Bus-rail | Real world | Given parameter | Timetable and vehicle scheduling | SA |

### 2.6. Stop spacing

"Stop spacing" is another major identified sub-category among published studies that have applied mathematical programming and optimization algorithms to study PBTNDPs&OP. Although the studies in this sub-category can also be considered as the published papers related to "Network Design", "stop spacing" has been considered as an independent sub-category to have a closer and more precise look at these articles. Kikuchi (1985) conducted one of the earliest studies in this area. They used mathematical programming to investigate the relationships between the number of bus stops and service frequency. To this end, they developed an optimization model to joint optimization of the number of stops and service frequency for a single route transit network. They assumed that the fleet size is a given parameter, and found that under this assumption, unique optimal solutions can be obtained for the decision variables, leading to the minimum average travel time of the passengers. Spasovic and Schonfeld (1993) considered a radial transit network with many-to-one transit service and uniform demand pattern, then, developed a model to simultaneously optimize transit route length, stop location, route spacing, and service frequency with the objective of minimizing the total operation costs, waiting costs, and access costs. To handle the complexity of the proposed methodology, they also suggested an optimization algorithm and applied it to a rectangular and wedge-shaped transportation network. As the key findings, they showed that demand patterns would significantly affect the stop locations. Also in the optimized condition, there would be a balance between the different cost factors.

The impacts of demand patterns on bus stop locations have been analyzed by Ibeas et al. (2010). Using bi-level programming, they investigated how demand density function, service frequency, in-vehicle crowding, private car traffic flow, the socio-demographic characteristics of each service area, the available budget of the agency, and fleet size could impact the optimal location of the bus stops. The objective function of the proposed model was to minimize the total social costs in the network. Alonso et al. (2011) investigated the relationship between the congestion level in the city and optimal bus stop location. They developed bi-level programming, with the upper level responsible for determining optimal stop locations under various levels of congestion in order to minimize total social costs in the network, and the lower level being a mode-split assignment model. Applying the proposed model to the transit network of Santander City, Spain, they showed that for high demand, bus stops will be located closer to each other, and as a result, the accessibility to the public transit service would be increased.

Combining a GIS with mathematical programming, Delmelle et al. (2012) studied the conflict between accessibility by having more bus stops and shorter in-vehicle travel time. They also considered the stops' attractiveness for passengers and proposed a coverage model for this problem. To cope with computational complexities, they also suggested a SA algorithm and applied it to a test network. Moura et al. (2012) used two-stage optimization programming to obtain the optimal bus stop locations. In the first step, they suggested a macroscopic model for this problem with the objective of minimizing the total social costs in the network. Then, they proposed a microscopic model to obtain bus locations using the results of the first stage and considering the maximization of the bus's speed as the objective function. They tested the performance of this two-stage process to solve the bus stop location problem for a real transit network under different scenarios for signal schedule, traffic, and bus flow. In addition to the trade-off between accessibility costs and travel costs, Ceder et al. (2015) considered the impacts of uneven topography on the optimal bus locations. They assumed that uneven topography can affect the walking cost, public bus access routes' attractiveness, and in-stop acceleration rates. Having these assumptions, they proposed a mathematical model and a heuristic algorithm to deal with bus stop location problems and applied them to the PBTS of Auckland, New Zealand. Chen et al. (2018) provided bi-objective programming for bus stop location problems, in which minimizing bus dwell time was one of the objectives, and bus stop number minimization was another objective function. The network's service accessibility, traffic congestion, and traffic flow were also considered in the modeling. Applying the model to the transit network of Yancheng, China, they showed that the model is able to find Pareto solutions for real-size networks using CPLEX.

Luo et al. (2020) integrated continuum approximations with nonlinear mathematical programming to study the bus stop location problem. They used the results of continuum approximations as the inputs for the proposed non-linear optimization model. First, using continuum approximations, they obtained the optimal bus route design. Then, based on the optimal design of transit routes in the first step, they applied a mathematical programming to obtain the optimal location of the stops. They also suggested an interior-point algorithm to find near to optimal solutions for the large size networks and tested it with a group of numerical instances and a real case of Chengdu, China. Zhang et al. (2020) studied the bus stop location problem for an on-demand transit system. To this end, they developed mathematical programming with the objective of minimizing the total travel time with respect to bus size, route characteristics, and the minimum allowed distance between each consecutive stop. Solving different numerical examples, they showed that the locations of the bus stops for on-demand service and fixed-route services will be totally different. They discovered that demand density, access, in-vehicle, and waiting costs will all have a significant impact on the results for both systems. Taplin and Sun (2020) focused on the bus stop location and bus routing problems for a feeder bus transit. They first found the optimal stop locations, aiming to find the minimum access cost. Then, they used a GA to evaluate the candidate stop locations, in order to maximize the demand for public transit. To solve the routing problem, a traveling salesman GA was used in the last step. They claimed that to deal with this problem, the agency must consider both accessibility and route length in the planning step to achieve more reliable results.

Table 6 represents the main features of the reviewed papers in this subsection. According to this table, access time, waiting time, in-vehicle time, and operation costs are the main cost factors, and stop location is the main decision variable considered by the reviewed papers in this section. The considered constraints in the proposed mathematical programming are mainly related to the fleet size and vehicle capacity. In this section, different optimization algorithms have been applied to solve the proposed models.

**Table 6.** Main features of the reviewed papers in subcategory 6.

| Reference | Objective Function | Constraints | Transit mode | Network structure | Demand Pattern | Decision Variables | Solution Approach |
|---|---|---|---|---|---|---|---|
| Kikuchi (1985) | Min. sum of access, waiting and in-vehicle times | Fleet size | Bus | - | Given parameter | Stop spacing, service frequency | Graphical |
| Spasovic and Schonfeld (1993) | Min. sum of access, waiting, operational cost | Vehicle capacity | Bus | Radial transit network | Uniform many-to-one demand pattern | Route length, route spacing, headway, stop spacing | A penalty based algorithm |
| Ibeas et al. (2010) | Min. sum of waiting, in-vehicle, access, transfer and operational costs | Max agency costs, fleet size, max/min service frequency | Bus | Real world | Demand function | Stop spacing | Hook-Jeeves algorithm, ESTRAUS |
| Alonso et al. (2011) | Min. sum of operational cost, access, waiting, in-vehicle, transfer costs | Fleet size | Bus | Real world | Elastic demand | Stop location | Tabu search algorithm |
| Delmelle et al. (2012) | Max. interaction between all demand nodes and facilities | Interactions, maximum number of selected facilities | Bus | Real world | Given parameter | Stop location | SA |
| Moura et al. (2012) | Max. commercial speed of the public transport service, min. sum of operational cost, access, waiting, in-vehicle, transfer costs | Fleet size | Bus | Real world | Given parameter | Stop location | Tabu search algorithm |
| Ceder et al. (2015) | Min. sum of cruising time, dwell time, acceleration time and layover time, personnel costs, idle costs, min. total access time, min sum of total access time, cruising time, dwell time, acceleration time | Stop-demand matching | Bus | Real world | A continuously varying function along a route and originate from a discrete set of points | Stop location | A heuristic evolutionary algorithm |
| Chen et al. (2018) | Min. total dwell time at stops, Min. total number of bus stops | Accessibility level | Bus | Real world | - | Stop location | CPLEX |
| Luo et al. (2020) | Min sum of waiting, access in-vehicle & operational costs | Restricted locations, stop capacity, bus capacity | Bus | Real world | Location- and time-dependent density function | Stop location | MATLAB: fmincon |
| Zhang et al. (2020) | Min. total travel time | Segment length, shed line configuration, minimum stop spacing, vehicle capacity | Bus | Example network | Density function | Stop spacing | Mathematica: dynamic programming function |
| Taplin and Sun (2020) | Max. walking interaction between dwelling and stop, min. distance (shortest path) | - | Bus | Real world | A function | Routing, stop spacing | GA |

## 2.7. Mode split and traffic assignment

Model split and traffic assignment for multimodal transit networks, where bus transit is one of the modes, are other identified sub-problems. In most of the studies in this sub-category, mode choice, route choice, or traffic assignment problems are being solved as a part of another problem such as routing. In this case, most of the problems are modeled as bi-level programming, where the traffic assignment problem has been solved at the lower level.

LeBlanc (1988) conducted one of the pioneer studies in this area. They proposed a mathematical programming model to jointly optimize service frequency associated with each transit route and mode-split assignment in the network. They analyzed the impacts of mode split on the service frequency changes in each route and tried to provide more precise estimations of users' mode choice and flows on each link in the network by considering accessibility, in-vehicle time, and delays caused by transfers. They used the Hooke-Jeeves algorithm to cope with the complexity of the proposed methodology. Spiess and Florian (1989) studied the transit assignment problem for a network with fixed transit routes, assuming that users will choose a mode that leads to the minimum expected travel time. Ignoring congestion effects, they proposed a linear optimization model supposing that waiting time in each stop will be just a function of combined frequency. The complexity of the proposed model was increased linearly by an increase in the network size. They extended the proposed model to the case of non-linear cost and proposed a label setting algorithm to handle the polynomial-time issue.

Shih et al. (1998) developed a mathematical programming model to study the joint trip assignment, transit network design, and coordinated operations planning problems. The developed methodology was also used to obtain the optimal bus sizes. The possibility of using on-demand transit systems to serve passengers who were not served by fixed-route transit systems during the peak hours was considered in their study. Their methodology includes four sections: bus transit route generations, traffic assignment, bus capacity and service frequency determination for both coordinated and non-coordinated operations, evaluating and selecting a transit center for the coordinated operation, and finally improving the transit network's performance by improving the generated transit routes. The proposed methodology was validated through the bus transit network of Austin, Texas. Gao et al. (2004) developed a bi-level optimization model to deal with the joint transit network design and transit assignment problems. The upper-level model was responsible for designing an efficient transit network, while the lower level was dealing with the transit equilibrium assignment problem for the designed network. They proposed a sensitivity analysis-based heuristic algorithm to manage the complexity of the model and tested it with a numerical example.

Fan and Machemehl (2006) developed a multi-objective nonlinear mixed-integer programming to design an optimal bus transit network and integrated it with a GA to find near-optimal solutions for large scale transit networks, taking into account a variable travel demand pattern. The proposed methodology was used to generate a set of feasible transit routes, estimate the O-D matrix, obtain associated service frequency, and solve the traffic assignment problem. Solving the problem for a test network, they analyzed the performance of the proposed methodology. Shimamoto et al. (2010) proposed bi-level programming to solve the bus transit network design problem in the upper level and the traffic assignment problem in the second level. The upper-level model itself was also a bi-level programming with the objective of minimizing the total agency's and users' costs and analyzing the impacts of a reduction in the agency's costs on the users' costs. The goal of the proposed model was to evaluate an existing bus transit network to find possible solutions that lead to improvement in the network's performance. The model was applied to the bus transit network of Hiroshima City. The results for this city showed that the current system is working near the Pareto solutions. However, trying to decrease the costs of the agency will lead to higher costs and inequality for the users.

Mesbah et al. (2011b) provided a bi-level model to study the exclusive lane assignment problem for public buses. The model in the upper level was responsible for identifying the optimal set of exclusive bus lanes in the network, while the model in the lower level was related to the modal split, traffic assignment, and transit assignment problems. To cope with the complexity of the proposed bi-level programming, a decomposition methodology was suggested that was able to find near to optimal solutions for the mentioned problem. The performance of this methodology was verified through numerical examples. In another study, Mesbah et al. (2011a) suggested a bi-level optimization model similar to the previous study with the objective of reassigning road space in the transportation network to public buses and private cars. The model was used to solve the modal split, user equilibrium traffic assignment, and transit assignment problems, while the objective function in the upper level was to minimize the agency's costs and in the lower level it was to minimize the users' costs. A GA is suggested to solve the model to find an acceptable solution within an acceptable running time and tested on an example network.

Considering a bi-modal bus-private car transportation network, Yu et al. (2015) developed a bi-level optimization model to study the bus lanes network design problem. The proposed model in the upper level was responsible for finding the optimal service frequency and lane combination, leading to the minimum average travel time and the existing difference among users' comfort in different bus lanes. The mode split and traffic assignment problems are modeled on the lower level. The column generation algorithm, the B&B algorithm, and the MSA are integrated to solve the proposed method and were applied to a test example based on the transit network in Dalian. Considering a stochastic demand pattern in a public transit network with two services, including a rapid fixed system and a flexible transit route system, An and Lo (2015) studied the bi-modal transit network design considering user equilibrium traffic assignment assumption by providing robust mathematical modeling. They took urban development density into account in their analysis to find the minimum density needed to make the rapid fixed transit system economically sustainable. The model was converted to a mixed-integer linear optimization model and integrated with a cutting constraint algorithm to solve the problem for the real scale networks. Using mathematical modeling, Cao and Wang (2017) studied the optimal user assignment problem in the customized bus transit service with the objective of minimizing in-vehicle waiting costs, delay costs, and operation costs. A B&B based algorithm was also suggested to solve the proposed model and, was verified through a numerical example. The results showed that, compared to regular buses, having customized buses in the network will lead to a more cost-efficient transit system.

Liu et al. (2018) considered a transit network where park-and-ride sites are located in the suburban areas, and a rapid bus system that transfers passengers between the sites and rail transit stations. They used convex mathematical programming to solve the modal split and traffic assignment problems for this network to analyze the impacts of the suburban park-and-ride sites on the traffic flow in the network. The Evans and self-adaptive gradient projection algorithms are integrated to handle the computational complexities. Solving a group of numerical examples, they showed that having park-and-ride sites in the suburbs can lead to significant benefits for the transit network, such as public transit service promotion. Considering stochasticity in the transit network and service reliability, Wu et al. (2019) proposed a bi-level optimization model to study the joint bus schedule coordination and users' route choice/demand assignment problems. They assumed that in the case of missing a connection, users would have two strategies, modifying their route or adhering to an already selected route. Using a mixed-integer non-linear optimization model in the upper bound, the service schedule coordination problem was solved with the objective of minimizing the total users' and agency costs. In the lower level, an optimization model was related to the route choice problem. Finally, a heuristic was integrated with SAM to find near to optimal solutions at an acceptable execution time.

Table 7 represents the main features of the reviewed papers in this subsection.

**Table 7.** Main features of the reviewed papers in subcategory 7.

| Reference | Objective Function | Constraints | Transit mode | Network structure | Demand Pattern | Decision Variables | Solution Approach |
|---|---|---|---|---|---|---|---|
| LeBlanc (1988) | Min. sum of total auto trips, and frequencies | Min service frequency | General | Sioux Falls network | - | Service frequency, mode-split assignment | Dial's transit loader to solve Frank-Wolfe sub-problems, Hooke-Jeeves algorithm |
| Spiess and Florian (1989) | - | - | General | An example | Total demand is known | Trip assignment | A label-setting algorithm |
| Shih et al. (1998) | - | - | Bus | Real world | - | Trip assignment, vehicle size, routing, service frequency | Heuristic |
| Gao et al. (2004) | Min. the total deterrence of transit system and cost caused by frequency setting | Flow conservation | General | An example | Given O-D demand matrix | Transit equilibrium assignment, service frequency | Heuristic |
| Fan and Machemehl (2006) | Min. sum of travel, transfer, operation and unsatisfied demand costs | Capacity & fleet size | Bus | Real world | Inelastic many-to-many | Routing & travel time between nodes | Heuristic |
| Shimamoto et al. (2010) | Min. sum of travel time, waiting time, failure to board cost & operational costs | Fleet size, max travel time on a line | Bus | Real world | Given O-D demand matrix | Transit assignment, routing, service frequency | NSGA-II |
| Mesbah et al. (2011b) | Min. sum of the total travel time emission, | Budget | Bus-car | Test network | Given O-D demand matrix | Exclusive bus lanes, modal split, user | Decomposition methodology |
| | noise, accident, and reliability of travel time by car and bus | | | | | equilibrium traffic assignment, and transit assignment problems | |
| Mesbah et al. (2011a) | Min. sum of the total travel time emission, noise, accident, and reliability of travel time by car and bus | Budget | Bus-car | Test network | Given O-D demand matrix | Transit priority, modal split, user equilibrium traffic assignment, transit assignment | GA |
| Yu et al. (2015) | Min. average travel time of travelers, min. difference of users' comfort among all the bus lines | - | Bus | Real world | Given O-D demand matrix | Bus lanes, service frequency, transit assignment, modal split | The column generation algorithm, B&B, the method of successive averages |
| An and Lo (2015) | Min. sum of operating costs, construction cost, station construction cost, and users travel time. | Rapid transit line constraints, variational inequality, max/min service frequency, link-line matching | Bi-modal: rail-bus | An example | Stochastic demand | User equilibrium, urban density, transit line construction sequence, service frequency | A cutting constraint algorithm |
| Cao and Wang (2017) | Min. sum of average travel time, average waiting time, average delay penalty, average ticket price | Fleet size, bus load, arrival time window | Customized Bus | An example | Given O-D demand matrix | User assignment | B&B, simulation |
| Liu et al. (2018) | - | Flow conservation, budget, capacity | Multi-modal: bus, train | A linear corridor, Sioux-Falls network | Given O-D demand matrix | Modal split, traffic assignment, the optimal location and capacity of RP&R sites | Exact: nonlinear valid inequalities, Evans algorithm, self-adaptive gradient projection |
| Wu et al. (2019) | Min. sum of in-vehicle, induced slack time, waiting, missed connection, delayed connection, delay and operational costs | Max/min headway | Bus | Medium-size hybrid bus network | Known and fixed | Bus schedule coordination (headways and slack times), demand assignment, rerouting of users | Heuristic, method of successive averages |

## 2.8. Network design and Dispatching policy

Jointly optimization of the network design and vehicle dispatching policies is one of the most popular and earliest problems in the area of PBTNDPs&OP. A large number of the identified studies are related to this subject. There are a number of review papers that have covered papers published before 2015, such as Ibarra-Rojas et al. (2015). Hence, for this subcategory also only the selected papers published after 2015 will be reviewed.

Schöbel (2017) studied the simultaneous transit routing, timetabling, and bus scheduling optimization problems, providing multi-objective programming. In addition to the optimization model, they also suggested an eigenmodel to investigate the mentioned problem and discussed how this method can be used to develop a heuristic algorithm to study this problem. The quality of the proposed algorithm from different perspectives was analyzed through numerical examples. In order to design an optimal timetable for a public transit network, Laporte et al. (2017) considered the route choice behavior of the users. They assumed that timetabling would impact the routing behavior of the users and provided a multi-objective optimization model to solve the timetabling problem and also users assignment to the best timetables. The objectives were users' costs minimization and the agency's costs minimization, and a constraint related to the vehicle capacity was considered in the proposed model. Focusing on customized bus transit network design problems, Tong et al. (2017) investigated the problem of finding the optimal transit routes and service schedules with respect to the users' pickup-drop-up locations, their preferred service time window, and users-transit route matches. Their objective was to know how to keep the system profitable in the long term by meeting the minimum needed loading rate and maximizing the number of loaded users per bus. To this end, they provided a multi-commodity network flow-based programming, and to handle the computational complexity of this model, they proposed a Lagrangian based optimization algorithm capable of solving the problem for real size networks. The performance of the algorithm was examined through numerical instances and real cases. The problems of bus transit routing and the service frequency set associated with each transit route have been studied by Chu (2018). They came up with a mixed-integer model to jointly optimize routing and scheduling, and then proposed a parallel B&B and a Branch-price & cut algorithm to solve this model. Both variable and constant service frequencies were considered in their study. The performance of the proposed algorithm was verified by solving a bunch of numerical examples.

Ruano-Daza et al. (2018) proposed a bi-level methodology based on the Multiobjective Global-Best Harmony Search algorithm for joint optimal BRT network design and dispatching policy optimization problems. The algorithm was used to identify possible solutions, rank and evaluate the Pareto solutions with the multiple objectives of minimizing the unused vehicle capacities and the travel costs of the users. In the upper level, the algorithm was responsible for finding optimal routing strategies, while the lower level was related to the frequency setting problem for each solution of the upper level. Finally, they applied the proposed method to the real case of Pereira, Colombia, and compared the obtained results with the results of the NSGA-II and MOEA/D algorithms for the same problem. Buba and Lee (2018) tried to minimize the users' travel costs and unserved travel demands by designing optimal transit routes and associated dispatching policies for each designed route. To this end, they provided a differential evolution approach and applied it to the Mandl's Swiss network. The results showed that the provided method outperformed previously introduced methods. Owais and Osman (2018) first developed a modified multi-objective GA to find the optimal or near to optimal transit route combinations; then, they suggested a bi-level optimization algorithm to find service frequency for each transit route. They minimized the users' and agency's costs. The proposed method was used to solve routing and scheduling problems for two real world transit networks.

Another study on joint bus routing and service frequency optimization problems has been conducted by Jha et al. (2019). They suggested a two-stage procedure to solve this problem, where in the first stage, a GA was used to generate a set of candidate transit routes, and in the second stage, multi-objective programming was used to obtain service frequency. Minimizing users' travel costs and minimizing the agency's operation costs were the objective functions of the second stage. To solve the proposed model in the second stage, a multi-objective particle swarm algorithm with multiple search strategies was suggested and applied to the transit network Mandl's Swiss network, and its performance was compared to the performance of NSGA-II and other methods in the literature. Considering an integrated metro-bus transit network with a stochastic travel time and a stochastic demand pattern, Liang et al. (2019a) suggested a two-stage optimization procedure to find the optimal bus transit network design and service frequency associated with each bus transit route. In the first stage, a column generation method was provided to produce a feasible set of candidate transit routes, and in the second stage, using a stochastic linear optimization model, the optimal dispatching policies and passenger path flow were obtained. A primal-dual online algorithm was presented to solve the second stage. The proposed two-stage approach got validation through a numerical example and a real transit network in Beijing, China. Considering conflicts between users and agency costs, Liang et al. (2020) developed a multi-objective mathematical programming to deal with the joint bus routing and service frequency optimization problem. They considered two different populations, and to solve the proposed multi-objective model, a cooperative evolutionary algorithm was also suggested to coevolve these populations along with the considered objectives. Solving the problem for Mandl's test network using the proposed procedure and some of the other existing methods in the literature, they showed that the proposed procedure would provide more efficient results.

Taking transfers costs and impacts of the congestions on the travel times into consideration, Chai and Liang (2020) developed a multi-objective programming and integrated it with an NSGA-II algorithm to simultaneously optimize the transit routes combination, service frequency, mode choice, and demand assignments. The objective function was to minimize of users' travel time and required fleet size. The quality of the results obtained by the proposed procedure was compared to the other existing methods through a real-world case study. Considering a transit network with multiple routes and multiple terminals, Ranjbari et al. (2020) provided a three-stage optimization procedure to design an efficient transit network. In the first stage, a set of possible candidate terminals was identified, while in the second stage, a k-shortest-path algorithm was applied to find transit routes between identified terminals. In the final stage, a mixed-integer linear optimization model was suggested to obtain the optimal terminals and depot locations, route configuration, dispatching policies, and the required number of buses leading to the minimum users' travel time and bus deadheading time.

The demand-responsive feeder bus transit network design and service frequency setting problems have been studied by Lu and Wang (2020). They developed a compatibility-based algorithm to solve the mentioned problem, in which the objective was to achieve optimal values while minimizing operation costs and users' inconvenience. To make the results more realistic, they considered the time window for both the start of the travel and the end of the travel, the available fleet size, and the flexibility of the feeder transit services. The performance of the proposed method was evaluated through a group of numerical examples. Ning et al. (2021) developed a multi-objective programming to study bus routing and service frequency setting problems with the objectives of the number of served users maximization, the route lengths minimization, and the fleet size minimization. The users' satisfaction was also considered in the modeling process. For user flow distribution during periods with a similar pattern and a real-time user flow distribution, an offline algorithm and an online algorithm were presented, respectively, and were tested in a real case study.

Table 8 represents the main features of the reviewed papers in this subsection. According to this table, waiting time, in-vehicle time, transfer costs, users' inconvenience, and operation costs are the main cost factors considered by reviewed papers in this section. Also, routing and service frequency are the main optimized decision variables. The most commonly considered constraints are related to the available fleet size, network connectivity, max/min headway, allowable bus route length, and flow conservation.

**Table 8.** Main features of the reviewed papers in subcategory 8.

| Reference | Objective Function | Constraints | Transit mode | Network structure | Demand Pattern | Decision Variables | Solution Approach |
|---|---|---|---|---|---|---|---|
| Schöbel (2017) | Min. operation costs, min. sum of travel time and number of transfers | Max/min service frequency | Bus | An example | Given | Routing, timetabling, bus scheduling | Heuristic |
| Laporte et al. (2017) | Min. the total user inconvenience, min. the line runs costs & min. the fleet size costs | Vehicle capacity | General | An example | A fully disaggregated demand. | Timetabling, users assignment to the timetables | An ϵ-constraint solution |
| Tong et al. (2017) | Min. the number of unserved passengers, and the routing costs | Users' pickup-drop-up locations, users' preferred service time window, user-transit route matches | Customized bus | Real world | Real data | Routing, service schedules | Lagrangian based algorithm |
| Chu (2018) | Min. the weighted sum of bus operating cost, passenger generalized cost, and penalty for unsatisfied demand | Constant and variable headway, fleet size, dynamic and static travel time, flow conservation, connectivity and subtours | Bus | An example | Time-dependent O-D pairs | Routing, service schedules | Parallel B&B, Branch-price & cut algorithm |
| Ruano-Daza et al. (2018) | Min. the unused vehicle capacities and the travel costs of the users. | - | BRT | Real world | - | Routing, service schedules | Global-Best Harmony Search algorithm, NSGA-II, MOEA/D |
| Buba and Lee (2018) | Min. the users' travel costs and unserved travel demands | Min frequency, maximum flow occurring on any link, fleet size, maximum number of routes allowed in the given route set | Bus | Benchmark network | Given O-D matrix | Routing, service schedules | A differential evolution algorithm |
| Owais and Osman (2018) | Min. sum of the users' cost and agency's costs | Min direct demand coverage, maximum allowable bus route length, load factor, Max/min number of resulted network routes | Bus | Real world | Given O-D matrix | Routing, service schedules | Multi-objective GA |
| Jha et al. (2019) | Min. sum of users' travel costs & Min. the agency's operation costs | Fleet size, emissions, max/min frequency, route redundancy, cycle prevention, connectivity | Bus | Benchmark network | Given O-D demand matrix | Routing, service schedules | MOPSO, NSGA-II |
| Liang et al. (2019a) | Min. the sum of passenger cost, operator cost, and overflow penalty | Vehicle capacity, users flow assignment, arc capacity | Metro-bus | Real world | Stochastic demand pattern | Routing, service schedules, users path flow | Column generation, a primal-dual online algorithm |
| Liang et al. (2020) | Min. the total in-vehicle travel time, total waiting time and total number of transfers, & min. operational costs | Min/max frequency, min/max number of nodes on each route, vehicle capacity | Bus | Benchmark network | Heterogeneous demand | Routing, service schedules | A cooperative evolutionary algorithm |
| Chai and Liang (2020) | Min. users' travel time & min. required fleet size | Connectivity, transfer time, flow conservation, station capacity, min headway | Bus | Real world | Given O-D demand matrix | Transit routes combination, service frequency, mode choice, and demand assignments | NSGA-II |
| Ranjbari et al. (2020) | Min. users' travel time and bus deadheading time | Min. fraction of the total satisfied transit demand | Real world | Bus | Given O-D demand matrix | Routing, the optimal terminals and depot locations, route | K-shortest-path algorithm |
| | | | | | | configuration, dispatching policies, and the required number of the buses | |
| Lu and Wang (2020) | Min. operation costs and users' inconvenience | Time window for both starts of the travel and end of the travel, available fleet size, the flexibility of the feeder transit services | An example | Bus | Given data | Routing, service scheduling | A compatibility-based algorithm |
| Ning et al. (2021) | Max. the number of the served users, Min. the routes length, Min. the fleet size | Load factor, max waiting time | Real world | Shared bus | Dynamic real-time passenger flows | Routing, service scheduling | An offline algorithm, an online algorithm |

## 2.9. Network design and routing

Another popular and critical problem that has been studied in many research using mathematical programming and optimization algorithms is the design of the topology of bus transit networks and bus transit routing problems. A large portion of the identified published papers is related to this subcategory. However, as mentioned before, only published papers after 2015 will be reviewed in this section. For this subcategory, only papers focusing solely on routing or network topology design issues have been considered. Studies related to the integrated network design and dispatching policy problems have been reviewed in section 2.8.

Chen and Nie (2017) focused on the optimal hybrid fixed and demand adaptive flexible public transit network design problem. First, they assumed that pairing a flexible route with a fixed route would decrease the complexity of the problem, and then they developed a mathematical programming with the objective of minimizing total agency's and users' costs to jointly design both systems. To solve the proposed mixed integer optimization model, they used a commercial metaheuristic algorithm. The results indicated that paired flexible- and fixed- route systems outperform both sole flexible- and sole fixed-route systems. Fan et al. (2018) considered a bi-modal public transit service in which users are served by local bus transit and a mass express transit service such as rail transit. They assumed that these two transit networks would intersect each other through different locations in a city with a grid network structure to serve demand in different areas. Having this assumption, they developed a bi-level optimization model to jointly design the mentioned bi-modal transit network, taking the route choice behavior of the users into account in the modeling process. The objective function of the proposed model was to minimize total agency and users' costs. While the upper level model was related to the network design problem, the lower level was a traffic assignment model. Finally, they suggested an optimization algorithm to solve the model. Feng et al. (2019) focused on the redesigning of the transit route layouts problem, aiming to minimize transfer costs in a public bus transit network. To this end, they provided a mathematical modeling and a GA to find near to optimal solutions for the developed model. The model and algorithm were applied to the bus network of a city in China. The obtained results proved the capability of the proposed methodology to solve bus route design problems for large scale networks efficiently.

Ahmed et al. (2019) compared the performance of the different hyper-heuristics algorithms in the solving bus transit route design problem. The objective was to design an optimal bus transit network with the minimum users' travel time and the agency's operation costs. Each algorithm was applied to solve a group of test networks and their performances were compared from different perspectives, including solution quality and evacuation time. The results showed an integrated sequence-based selection method and great deluge algorithm would outperform other algorithms. Zhao et al. (2019) provided a bi-level optimization model to cope with the optimal exclusive bus lane network design problem with the minimum users' travel cost. They considered both public buses and private vehicle traffic in the modeling and suggested a GA algorithm to handle the complexity of the proposed model. Integrating GA with EMME, they solved some numerical and empirical instances and showed that consideration of the intersection operational dynamics in exclusive bus lane assignment would impact the results to a great extent. To deal with the uncertainty in the customized bus routing problem, Zhang et al. (2021a) proposed an uncertain mathematical modeling for this problem where the goal was to have an optimal route with the minimum total bus operation mileage. A modified GA was also suggested to handle the complexity and was applied to a case study to get validated. As a key result, they showed that considering uncertainty in the customized bus routing problem would lead to reduced operation mileage. Huang et al. (2021) studied the joint optimal exclusive bus lane network design and stop-skipping control planning problems. To this end, they developed a multi-step stochastic optimization model that was able to capture the uncertainty in the users' travel demand using a scenario tree. The complexity of the proposed approach for large scale problems was dealt with by converting them to small scale sub-problems. The performance of the provided methodology was tested through numerical examples.

Table 9 represents the main features of the reviewed papers in this subsection. According to this table, access costs, waiting costs, in-vehicle costs, transfer costs, and operation costs are the considered costs factors in the objective function of proposed models related to this section. Also, number of transit lines, route spacing, routing, and exclusive bus lanes are the main considered decision variables. Different optimization algorithms have been used to solve the proposed models. However, most of the identified papers related to this section have applied GA to find near to optimal solutions for their proposed models.

**Table 9.** Main features of the reviewed papers in subcategory 9.

| Reference | Objective Function | Constraints | Transit mode | Network structure | Demand Pattern | Decision Variables | Solution Approach |
|---|---|---|---|---|---|---|---|
| Chen and Nie (2017) | Min. sum of operation costs, walking, waiting, in-vehicle and transfer costs | - | Bus (fixed-flexible) | Grid network | Homogeneous spatial Poisson process | Service frequency, maximum walking distance, number of lines | A commercial metaheuristic solver, simulation |
| Fan et al. (2018) | Min. sum of operation costs, walking, waiting, in-vehicle and transfer costs | Vehicle capacity, min headway, spacing | Bimodal: Bus-rail | Grid network | Exogenous and inelastic | Traffic flow assignment, stop spacing | Gradient-based search |
| Feng et al. (2019) | Min. sum of operation costs, waiting time, walking time, transfer time & in-vehicle costs | - | Bud | Real world | Fixed given demand | Routing | GA |
| Ahmed et al. (2019) | Min. the passengers' travel time, and the operator's costs | - | Bus | Benchmark network | Given O-D matrix | Routing | Hyper – heuristics, a sequence-based selection method |
| Zhao et al. (2019) | Min. total users travel time | Flow conservation, lane assignment, max/min signal cycle length | Bus | Real world | Given parameter | Exclusive bus lane, signal timing | GA-EMME |
| Zhang et al. (2021a) | Min. the total mileage of vehicle operation | Time window for service receiving, rated user number in bus, number of bus stops | Customized bus | Real world | Uncertain demand | Routing | GA |
| Huang et al. (2021) | Min. sum of users waiting time, in-vehicle time, total trip time, total travel cots of car users, minimum total travel cost in the network | Skip-stop by two consecutive buses, skipping two contiguous stops by a bus | Bus-car | Benchmark network | Uncertain demand | Stop-skipping strategies, bus lane reservation, car flow, mode choice | A progressive hedging-based method |

## 2.10. Green and sustainable public bus network design

For this part, please see Mahmoudi et al. (2025).

## 2.11. Travel time and Reliability

Service reliability and travel time are important measures that affect the attractiveness of a PBTSs. There are a large number of studies that have focused on evaluating the reliability of an existing PBTS or improving travel times through PBTSs in the transit networks. However, most of these studies have used empirical methods and a small portion have applied mathematical programming. Xuan et al. (2011) integrated analytical approaches and mathematical programming to provide optimal dynamic bus holding strategies in order to help buses adhere to the schedule or operate with minimum slack time. They obtained closed form solutions for the single parameter case and near to optimal solutions for the other cases. Their results showed that transit systems designed under the proposed model would need almost 40% less slack time compared to the other methodologies. In order to analyze the bus dwell time, Meng and Qu (2013) developed a probabilistic model that was able to capture the between-bus interactions, users' arrival times, and shoulder lane traffic. To estimate the mean of the dwell time of the buses, they suggested a tangible procedure and applied it to a case study. Gkiotsalitis and Maslekar (2015) provided an integrated stochastic search and hopping/merging algorithm in order to find an optimal service schedule for the PBTSs leading to the minimum expected waiting time of the users. The main idea of the proposed methodology was to balance service headway variations by considering time intervals for bus dwell times during the service scheduling. The method could be used in both off-line and on-line versions. Applying this method to a real case in Asia, they showed that about a 50% improvement in the expected waiting time can be achieved by less running time and larger solution space searches.

Considering uncertainty in travel demand, An and Lo (2016) proposed a reliability based optimization model for the BRT network design problem. In addition to the BRT service, they also considered a flexible transit service that was responsible for serving passengers that were not served by the BRT because of capacity limitations. They developed a two stage optimization process, where transit routes and service frequencies were obtained in the first stage considering a predefined service reliability level, and in the second stage, the operation planning for the flexible service was obtained. They optimized the service reliability by minimizing the total costs of the BRT service, the flexible service, and the users. Finally, they integrated the gradient method and neighborhood search algorithm to handle the complexity of the proposed model and applied them to a group of examples. Gkiotsalitis and Cats (2018) studied the problem of designing a reliable dispatching policy for a PBTS by taking the variations in the travel demands, headways, and travel times into account. The objective was to obtain optimal service headway with respect to the bus size, fleet size, and agency's costs. They proposed a mathematical model for this problem and a Branch and Bound (B&B) to solve it. Solving some numerical examples, the proposed solving algorithm was show to have better performance than other heuristic algorithms.

Considering the variations in both travel time and bus dwell time, Gkiotsalitis et al. (2019a) focused on the timetable synchronization problem in a multi-route bus transit network. They developed a flexible minimax programming that was also capable of handling the problem of infeasibility. They validated the performance of the proposed model by applying it to the bus transit network of The Hague, Netherlands. Andrade-Michel et al. (2021) provided a mathematical programming to jointly optimization of drivers scheduling and vehicle scheduling with the objective of maximizing the users' satisfaction. The performance of the developed exact model was compared to the performance of a VNS algorithm by applying them to a group of numerical examples under different scenarios designed by Monte Carlo process. Liang et al. (2021b) investigated the bus holding problem using a non-linear programming that was converted to a one-dimensional line search model. An optimization algorithm was suggested based on the proposed models and was validated through solving numerical instances. As a key finding, they showed that in the case of having high travel demand in the transit network, bus bunching problems can be solved by the network itself, without the need to external controls.

Table 10 represents the main features of the reviewed papers in this subsection. Based on this table, different objective functions have been included in the proposed models in this section. However, as it can be seen from the considered decision variables, the goal of all published studies was to optimize dispatching policy in order to improve the reliability of the network.

**Table 10.** Main features of the reviewed papers in subcategory 12.

| Reference | Objective Function | Constraints | Transit mode | Network structure | Demand Pattern | Decision Variables | Solution Approach |
|---|---|---|---|---|---|---|---|
| Gkiotsalitis and Maslekar (2015) | Min. expected waiting time | - | Bus | Real world | Real data | Service scheduling | Stochastic search and branch hopping/ merging algorithm |
| An and Lo (2016) | Min. sum of link operation and construction costs, station construction cost, users in-vehicle, transfer & waiting costs & Min. total system costs | Station-node matching, connectivity, max/min service frequency & user flow conservation | General | Test network | Uncertain demand | Service reliability, transit line alignment & service frequency | The gradient method & neighborhood search |
| Gkiotsalitis and Cats (2018) | Min. sum of waiting, additional bus, operation and reliability costs | Fleet size & vehicle capacity | Bus | Real world | Spatial and temporal varying demand pattern | Service frequency | B&B and SQP |
| Gkiotsalitis et al. (2019a) | Minimax. travel time, dwell time & operation costs | Regulatory Constraints, max/min travel time & max/min dwell time | Bus | Real world | Real data | Service frequency & transfer synchronization | Relaxation |
| Andrade-Michel et al. (2021) | Min. sum of drivers, vehicles & uncovered demands costs | Trips overlap, drivers breaks overlap, drivers shift time, duty length, extra working hours, & trip-driver, trip-vehicle, and driver-vehicle compatibilities | Bus | Real world | Given set | Vehicle and driver scheduling | Exact, variable neighborhood search & simulation |
| Liang et al. (2021b) | - | - | Bus | Test network | Given parameter | Holding time | Line search method |

### 2.12. Disaster and disruption management

One of the most important problems in any transit network design problem is to design a robust transit system that can resist any unpredicted crisis, disaster, a disruption in the network. There are a small number of studies that have used a mathematical programming to address this problem for PBTSs. Sayyady and Eksioglu (2010) proposed a mixed-integer linear optimization model to find optimal public transit routes to manage evacuation during an unpredicted disaster. The objective of the study was to find optimal evacuation routes that can be used by public transit to serve the passengers that do not have access to private vehicles, with the objective of minimizing the total time of evacuation and the total number of losses. A Tabu search algorithm and a traffic simulation package were integrated to solve the proposed model for the transit network of Fort Worth, Texas. In another similar study, Swamy et al. (2017) proposed a multistage optimization model to find the best evacuation strategy prior to the hurricane using public transit services. To this end, they assumed that the fleet size, the locations of the shelters, the start time of disasters and evacuation zones were known. The proposed approach was used to find pickup locations, routing, allocating the demand areas to the shelters, and dispatching policies. They suggested a heuristic algorithm to solve the model. Liang et al. (2019b) focused on the problem of serving demand for public transit services using PBTSs when a disruption occurs in the rail transit system. To this end, they provided a model to design a robust path-based bus bridging service and suggested a column generation procedure to deal with the complexity of the proposed model and find near to optimal solutions. The proposed model and algorithm were applied to a case study, and the results were discussed. In another similar study, Tan et al. (2020) developed an evacuation optimization model using PBTSs during a disruption in metro services. The objective of the proposed model was to design a system with the minimum total cost of metro users by finding optimal bus transit routes and associated service frequencies. They took the uncertainty in the recovery time and different risk taking levels of the metro users into account in the modeling process and provided a heuristic algorithm to solve the model.

Mutlu et al. (2021) considered the COVID-19 outbreak and proposed a bi-level optimization model to find the optimal dispatching policy leading to the minimum virus transmission at the bus stops. To deal with the complexity of the proposed model, a differential evolution algorithm was suggested and was applied to solve the model for a small- and medium-sized hypnotically network. In another similar study, Chen et al. (2021a) developed a dynamic nonlinear integer optimization model to optimize boarding and alighting strategies in customized bus transit services with the objective of minimizing number of between-passengers contacts to control virus transmission during COVID-19 outbreak. Solving the model, they found multiple Pareto solutions for the problem where the high number of customized buses leads to a low number of passengers' contact and vice versa. Lu et al. (2021) investigated the evacuation problem using PBTSs where the objective was to have a minimum total evacuation time. They developed a mathematical programming to find optimal pickup nodes and optimal bus allocation, with respect to the constraints such as fleet size, network topology, and users and bus routes. The proposed model was used to solve the evacuation problem for the transit network of Sioux Falls, North Dakota.

Table 11 represents the main features of the reviewed papers in this subsection.

**Table 11.** Main features of the reviewed papers in subcategory 13.

| Reference | Objective Function | Constraints | Transit mode | Network structure | Demand Pattern | Decision Variables | Solution Approach |
|---|---|---|---|---|---|---|---|
| Sayyady and Eksioglu (2010) | Min. the total evacuation time and the number of casualties | Vehicle capacity, flow conservation & user boarding constraints | Bus-rail | Test network | Given parameters | Evacuation routes | Tabu search & simulation |
| Swamy et al. (2017) | Min. number of bus stops, Min total distance between pickup locations and shelters | Stops accessibility, shelter-stop matching & shelter capacity | Bus | Test network | Given parameters | Pickup locations, design transit route & dispatching policies | Heuristic algorithm - simulation |
| Liang et al. (2019b) | Min. sum of users in-bus and in-rail costs, demand overflow penalty & walking and transfer costs | Fleet size, rail users serving, start point of a trip, reserved capacity & max/min service frequency | Bus-Rail | Real world | Given parameters | Bus bridging lines, flow on each link, demand overflow & service frequency | A column generation procedure |
| Tan et al. (2020) | Min. the total cost of the affected metro passengers | Journey length | Bus-metro | Real world | Heterogeneous | Bus line & service frequency | A heuristic algorithm |
| Mutlu et al. (2021) | Min. the cumulative disease transmission risk cost & min. total travel time | Fleet size, max/min service frequency, | Bus | Test network | Given parameters | Service frequency & traffic assignment | Differential Evolution algorithm |
| Chen et al. (2021a) | Min. number of contact during COVID-19 outbreak | Number of users in the buses & user-vehicle-stop matching, | Customized bus | Real world | Real O-D matrix | Dispatching policy | Gurobi |
| Lu et al. (2021) | Min. evacuation during time | Fleet size, bus routes, pedestrian routes & network constraints | Bus | Real world | Given parameters | Pickup location & fleet allocation | Gams: MINLP solver |

## 2.13. Bus priority

Giving priority to public buses through priority lanes or transit signal priority (TSP) is a well-known efficient strategy to decrease travel time for public transit users and also increase the attractiveness of this service for the users. However, besides the technological side of this strategy, the operation planning side also needs to be investigated. Hadas and Nahum (2016) developed a multi-objective programming to find the optimal priority lane assignment for public buses. The objective functions were maximizing total travel time reduction, minimizing required budget, and balancing O-D terminals. In order to rank the obtained Pareto solutions, they also proposed a MCDM approach and obtained the proposed approaches to the transit network of Petah Tikva, Israel. Considering at-stop and in-vehicle users' delays as the evaluation criteria to investigate the priority level of TSP requests, Ye and Xu (2016) provided a mathematical model to give priority to the buses that have the same TSP requests. Considering a case study, they compared the performance of the proposed model with no-TSP cases, conventional TSP models, and recently proposed models, and showed that their model obtains way better results. Bingfeng et al. (2017) focused on the problem of the exclusive link assignment to the public buses and used a bi-level mathematical programming with the objective of minimizing total travel costs of the users to find the optimal set of the exclusive links. The mode choice and route choice behaviors of the travelers were also considered in the user equilibrium assignment problem, which was solved using a variational inequality model. To deal with the complexity of the proposed model, they suggested a B&B algorithm and applied it to solve a numerical example.

Anderson and Daganzo (2019) studied the problem of conditional TSP, where buses request priority only when it can lead to an improvement in service reliability, considering both scheduled and unscheduled PBTSs. To this end, they developed a mathematical model based on Brownian motion and integrated it with simulation to solve the problem. By limiting the number of priority requests, the results indicated that the traffic situation in the network would be better under conditional TSP compared to the traditional TSP. Taking a variable travel demand pattern, Ghaffari et al. (2020) proposed a bi-level robust programming to study the problem of part-time and full-time exclusive lane assignment to public buses. The non-linear optimization model at the upper level was responsible for identifying the optimal solution for the exclusive lanes leading to the minimum total travel time and hourly variation in the travel time in the network. The upper level model was a multi stage demand model. Zeng et al. (2020) tried to improve the service reliability of the PBTSs using TSPs and gathered data by CAVs. To this end, they proposed an online adaptive route-based TAP model and a local TSP model that, using them and real time data, the TSP operations can be updated continuously. They analyzed the performance of the proposed modeling using simulation studies.

Ghaffari et al. (2021) developed a bi-level programming to design a TSP plan under demand uncertainty. The objective function of the upper level model was minimizing expected social cost, and its constraints were related to travel time, available budget, and confidence level. The mode choice and traffic assignment problems were also solved at the lower level. To handle the complexity of the proposed model, an AC algorithm was suggested and used to solve the problem for a real network. In another similar study, Hao et al. (2021) focused on the TSP planning problem considering a stochastic travel demand pattern. The objective function of the proposed model in their study was the maximization of the total vehicular departure. An optimization search algorithm was also suggested based on the phase clearance reliability value and was applied to solve two case studies. Ren et al. (2021) proposed an optimization model to obtain optimal service scheduling for a BRT service with the objective of total travel time minimization. The proposed mixed integer nonlinear programming was used to jointly optimize the TSP schedule and bus stop schedules. A NSGA was suggested to find near to optimal solutions and was applied to solve the problem for a case study.

Table 12 represents the main features of the reviewed papers in this subsection. According to this table, the main decision variables in this section are exclusive bus lanes and priority strategies. Most of the reviewed papers consider an uncertain demand pattern to optimize these variables.

**Table 12.** Main features of the reviewed papers in subcategory 14.

| Reference | Objective Function | Constraints | Transit mode | Network structure | Demand Pattern | Decision Variables | Solution Approach |
|---|---|---|---|---|---|---|---|
| Hadas and Nahum (2016) | Max. total travel time savings, Min. construction costs & Max. minimal degree for all terminal nodes | Maximum alternative selection & path-alternative selection | Bus | Real world | Given parameters | Exclusive bus lanes, | SPEA2 |
| Ye and Xu (2016) | Min. sum of users in-bus delay & users waiting delay | TSP request generation, min green time, max priority time & pre-evaluation 0f a TSP request | Bus | Real world | Given | Priority level of a TSP request | Heuristic |
| Bingfeng et al. (2017) | Min. total users travel cost | Demand conservation | Bus | Small hypothetical network | Fixed given O-D demand matrix | Exclusive bus lanes | B&B |
| Anderson and Daganzo (2019) | - | - | Bus | - | - | Reliability & transit signal priority strategies | Simulation |
| Ghaffari et al. (2020) | Min. social costs | Travel time & budget | Bus | Real world | Uncertain demand | Mode choice, traffic assignment & exclusive bus lanes | Ant colony algorithm |
| Zeng et al. (2020) | Min. deviations from optimal timing at intersection(s) | Signal coordination, bus route trajectory, conversation of local time to system time & bus performance measures | Bus | Hypothetical network | Real time data | Signal timing | Simulation |
| Ghaffari et al. (2021) | Min. expected social cost | Total travel time, user satisfaction & budget | General | Real world | Uncertain demand | Mode choice, traffic assignment & exclusive bus lane | Ant colony algorithm |
| Hao et al. (2021) | Max. total vehicular departure | Green time ratio, outflow rate & queue clearness | Bus | Real world | Stochastic demand | Bus signal priority | A heuristic algorithm |
| Ren et al. (2021) | Min. the travel time of BRT bus | Signal controls, on-time rate of bus schedule, arrival time, intersection saturation degree, vehicle capacity & bus departure interval | BRT | Real world | Stochastic demand | Service scheduling | NSGA |

### 2.14. Emerging technologies and electric buses

For this part, please see Mahmoudi et al. (2025).

## 3. Possible extensions and future research directions for the applications of Mathematical programming and Algorithms in PBTNDP&OPs

Analyzing the provided literature reviews and tables for the second category, one can find existing research gaps and conduct new studies to cover these gaps. However, in this section, some possible extensions related to the applications of mathematical programming and optimization algorithms in PBTNDP&OPs have been discussed. In general, subjects that have not been yet analyzed or a few attempts have been made to consider them are discussed in the following (see Mahmoudi et al. (2024) and Mahmoudi et al. (2025) for identified research gaps related to emerging technologies and services and sustainable PBTND&OPs):

- **Complexity, algorithms and Hybrid methods**. Most of the developed models using mathematical programming and optimization algorithms for PBTNDP&OPs are highly complicated (e.g. NP-hard). Because of this complexity, most of the studies suggest an optimization algorithm (e.g. heuristics, metaheuristics, etc.) to handle this complexity and solve the proposed models for large-scale real-world networks. The literature review shows that GA is the most popular algorithm in this area because it fits well with proposed models for the PBTNDP&OPs. However, there are some concerns about this trend. First of all, GA and its modified forms can be considered as one of the oldest metaheuristics, having been introduced in the 1960s. Although most researchers in the field of PBTNDP&OPs use GA in their studies, it will be interesting to compare the results and performance of other algorithms, particularly recently introduced ones, to those of GA. In particular, knowing that the efficient performance of optimization algorithms is a critical factor in evaluating the quality of any optimization study and its results, the researchers should be encouraged to challenge themselves by handling the complexity of the developed models with more efficient algorithms. For more discussions about applications of metaheuristics in PTSs see Iliopoulou et al. (2019).

Secondly, the quality of the obtained near to optimal results and the execution time are two important factors in evaluating the performance of any optimization algorithm. Both of these factors are highly impacted by initial solutions and feasible set boundaries. An efficient strategy can be integrating analytical approaches by mathematical programming for studying the PBTNDP&OPs. Using analytical methods, a good initial solution for the optimization algorithms and also a set of boundaries for the different decision variables and other terms in the mathematical modeling can be obtained. Both of them can improve the quality of the performance of the proposed optimization algorithms and also the running time of computer programs. Hence, developing hybrid models for the PBTNDP&OPs based on analytical methods and mathematical programming can be another research direction that has not attracted much attention yet.

- **Cost/utility function**. Most of the developed models for the PBTNDP&OPs have considered cost/benefit factors based on homogeneous assumptions for the type of travel, type of passengers, type of the routes, type of the stations, and type of the transit modes. It means, for example, the waiting time factor for all passengers on all routes, all stops/stations, and all time periods will be a specific constant factor like a. But in the real world, the facts are totally different. For example, waiting costs for the juniors, seniors, employed, unemployed, students, men, and women will be different and also highly depend on the quality of the stop/station facilities, transit mode, and time of travel. Although considering these facts will lead to more realistic results for the PBTNDP&OPs, no study has developed mathematical modeling based on these assumptions. For more discussion about the impacts of service types and passenger types on the users' costs, see Ansari Esfeh et al. (2021).

## 4. Conclusions

PTSs are always assumed to be the most efficient and sustainable transit mode. In particular, when it comes to daily intercity communications, public bus transit services (PBTSs) are assumed to be one of the most popular transit services among city planners. As PBTS is one of the oldest public transit modes and, at the same time, one of the cheapest and easiest modes of public transit to add to an urban transportation network, it has been turned into the backbone of the transportation network of many small, medium, and large cities all around the world. PBTS, with its different versions (regular buses, BRT, feeder, shuttle, school bus, etc.), is applicable in any city with any transportation network size/structure and any population density. Because of this popularity and applicability, the public bus transit network design problem and operations planning (PBTNDP&OP) have attracted a lot of researchers' and transportation managers' attention. As a result, there are a good number of publications that have focused on different problems related to the PBTNDP&OP, using different methodologies. Mathematical programming is a popular methodology that researchers have widely applied to approach these problems.

In this paper, a comprehensive literature review has been conducted on the application of mathematical programming in PBTNDP&OPs. The paper includes two major sections: First, in 15 subcategories, the existing literature related to the applications of mathematical modeling in PBTNDP&OPs was reviewed. The identified subcategories for this method are: 1) Accessibility and coverage, 2) Feeder transit network design, 3) Service pricing and fare management, 4) Fleet management, 5) Dispatching policy, 6) Stop spacing, 7) Mode split and traffic assignment, 8) Network design and dispatching policy, 9) Network design and routing, 10) Green and sustainable public bus network design, 11) Flexible and fixed transit service, 12) Travel time and reliability; 13) Disaster and disruption management, 14) Bus priority, and 15) Emerging technologies. Second, after comprehensively reviewing the selected papers, the identified future directions and possible extensions were discussed in the paper.

The authors believe that the provided taxonomy, review, and research directions in this paper can highlight different extensions and gaps for future studies and will inspire new research on the applications of mathematical modeling in the PBTNDP&OPs.